\documentclass[12pt,amsmath,amssymb,aip,jcp,preprint,floatfix]{revtex4-1}
\usepackage{graphicx}
\usepackage{color}
\usepackage[version=3]{mhchem}
\usepackage{dcolumn}
\usepackage{multirow, makecell}
\usepackage[normalem]{ulem}
\usepackage{cmbright}
\usepackage[dvipsnames]{xcolor}
\usepackage{array}
\usepackage{siunitx}
\usepackage{textgreek}
\usepackage{csquotes}
\usepackage{longtable}
\usepackage{comment}

\begin{document}

\newcommand\bfec[1]{\textcolor{orange}{\textit{ XXX BFEC: #1 XXX }}}
\newcommand\jj[1]{\textcolor{cyan}{\textbf{ JJ: #1 }}}

\setlength{\tabcolsep}{12pt}

\DeclareFontFamily{OT1}{cmbr}{\hyphenchar\font45 }
\DeclareFontShape{OT1}{cmbr}{m}{n}{%
  <-9>cmbr8
  <9-10>cmbr9
  <10-17>cmbr10
  <17->cmbr17
}{}
\DeclareFontShape{OT1}{cmbr}{m}{sl}{%
  <-9>cmbrsl8
  <9-10>cmbrsl9
  <10-17>cmbrsl10
  <17->cmbrsl17
}{}
\DeclareFontShape{OT1}{cmbr}{m}{it}{%
  <->ssub*cmbr/m/sl
}{}
\DeclareFontShape{OT1}{cmbr}{b}{n}{%
  <->ssub*cmbr/bx/n
}{}
\DeclareFontShape{OT1}{cmbr}{bx}{n}{%
  <->cmbrbx10
}{}

\renewcommand{\rmdefault}{cmbr}
\renewcommand{\sfdefault}{cmbr}

\title{Predicting the photodynamics of cyclobutanone triggered by a laser pulse at 200 nm and its MeV-UED signals -- a trajectory surface hopping and XMS-CASPT2 perspective}

\author{Ji\v{r}\'{i} Jano\v{s}}%
\affiliation{Department of Physical Chemistry, University of Chemistry and Technology, Technická 5, Prague 6, 166 28, Czech Republic}%
\affiliation{Centre for Computational Chemistry, School of Chemistry, University of Bristol, Bristol BS8 1TS, United Kingdom}%

\author{Joao Pedro Figueira Nunes}%
\affiliation{Diamond Light Source Ltd, Didcot, UK.}%

\author{Daniel Hollas}
\affiliation{Centre for Computational Chemistry, School of Chemistry, University of Bristol, Bristol BS8 1TS, United Kingdom}%

\author{Petr Slavíček}%
\email{petr.slavicek@vscht.cz}
\affiliation{Department of Physical Chemistry, University of Chemistry and Technology, Technická 5, Prague 6, 166 28, Czech Republic}%

\author{Basile F. E. Curchod}
\email{basile.curchod@bristol.ac.uk}
\affiliation{Centre for Computational Chemistry, School of Chemistry, University of Bristol, Bristol BS8 1TS, United Kingdom}%

\date{\today}%

\begin{abstract}
This work is part of a prediction challenge that invited theoretical/computational chemists to predict the photochemistry of cyclobutanone in the gas phase, excited at 200~nm by a laser pulse, and the expected signal that will be recorded during a time-resolved megaelectronvolt ultrafast electron diffraction (MeV-UED). We present here our theoretical predictions based on a combination of trajectory surface hopping with XMS-CASPT2 (for the nonadiabatic molecular dynamics) and Born--Oppenheimer molecular dynamics with MP2 (for the athermal ground-state dynamics following internal conversion), coined (NA+BO)MD. The initial conditions were sampled from Born--Oppenheimer molecular dynamics coupled to a quantum thermostat. Our simulations indicate that the main photoproducts after 2~ps of dynamics are CO + cyclopropane (50\%), CO + propene (10\%), and ethene and ketene (34\%). The photoexcited cyclobutanone in its second excited electronic state S$_2$ can follow two pathways for its nonradiative decay: (\textit{i}) a ring-opening in S$_2$ and a subsequent rapid decay to the ground electronic state, where the photoproducts are formed, or (\textit{ii}) a transfer through a closed-ring conical intersection to S$_1$, where cyclobutanone ring opens and then funnels to the ground state. Lifetimes for the photoproduct and electronic populations were determined. We calculated a stationary MeV-UED signal [difference pair distribution function -- $\Delta$PDF$(r)$] for each (interpolated) pathway as well as a time-resolved signal [$\Delta$PDF$(r,t)$ and $\Delta I/I(s,t)$] for the full swarm of (NA+BO)MD trajectories. Furthermore, our analysis provides time-independent basis functions that can be used to fit the time-dependent experimental UED signals [both $\Delta$PDF$(r,t)$ and $\Delta I/I(s,t)$] and potentially recover the population of photoproducts. We also offer a detailed analysis of the limitations of our model and their potential impact on the predicted experimental signals. 
\end{abstract}

\maketitle

\section{Introduction}

The field of nonadiabatic molecular dynamics simulations, concerned with the dynamics of molecular systems beyond the Born--Oppenheimer approximation, has experienced a growing interest by the physical chemistry and chemical physics community over the past three decades.\cite{tully2012perspective,persico2014overview,agostini2019different,https://doi.org/10.1002/anie.201916381,gonzalez2020quantum} This interest was greatly stimulated by the development of frameworks for on-the-fly nonadiabatic dynamics around the 90s,\cite{Tully1990,Tully1998,martinez1996multi,martinez1997non,Bittner1995,kapral:8919,Donoso1998,coker1995methods} combined with advances in efficient electronic-structure methods for excited states,\cite{PhysRevA.26.2395,finley1998multi,KOCH199575,doltsinis02,casida95,petersilka96,gonzalez2012progress,Matsika2021} which allowed for dynamics simulations of photophysical and photochemical processes unraveled by the emergence of femtochemistry.\cite{doi:10.1021/jp001460h} The field of nonadiabatic dynamics has continued its development over the years and reached a level of maturity that permits its use to interpret complex spectroscopic experiments or rationalize photophysical and photochemical events. 

Nonadiabatic molecular dynamics remains a challenging technique to operate: the overall nuclear dynamics is sensitive to the photoexcitation process\cite{suchan2018importance,doi:10.1021/acs.jctc.3c00024,doi:10.1021/acs.jpca.3c02333} and is out-of-equilibrium in nature, accounting for nuclear quantum effects is often required,\cite{crespo2018recent,curchod2018ab,persico2014overview,subotnik2016understanding,doi:10.1098/rsta.2020.0377,doi:10.1021/acs.accounts.7b00220, Ghosh2020, Agostini2018}  and including the effect of an environment challenges most frequently used methods for this purpose.\cite{D0CP05907B,doi:10.1021/acs.jpca.2c04756,TAVERNELLI2011101} 
Perhaps most importantly, the nonadiabatic dynamics can be very sensitive to the electronic-structure method employed to describe the different potential energy surfaces and couplings between them.\cite{Janos2023} While photophysical processes, where a photoexcited molecule explores a limited portion of the nuclear configuration space, are often less problematic to simulate for larger molecular systems than photochemical processes, the formation of photoproducts stresses more the approximations underlying nonadiabatic molecular dynamics simulations. Hence, it appears fair to say that, in 2024, nonadiabatic molecular dynamics simulations can be performed to study the photochemistry of small- to medium-size molecules, \textit{in vacuo} preferably, but at the cost of extensive tests prior to any production runs.

\begin{figure}[ht!]
    \centering
    \includegraphics[width=0.8\textwidth]{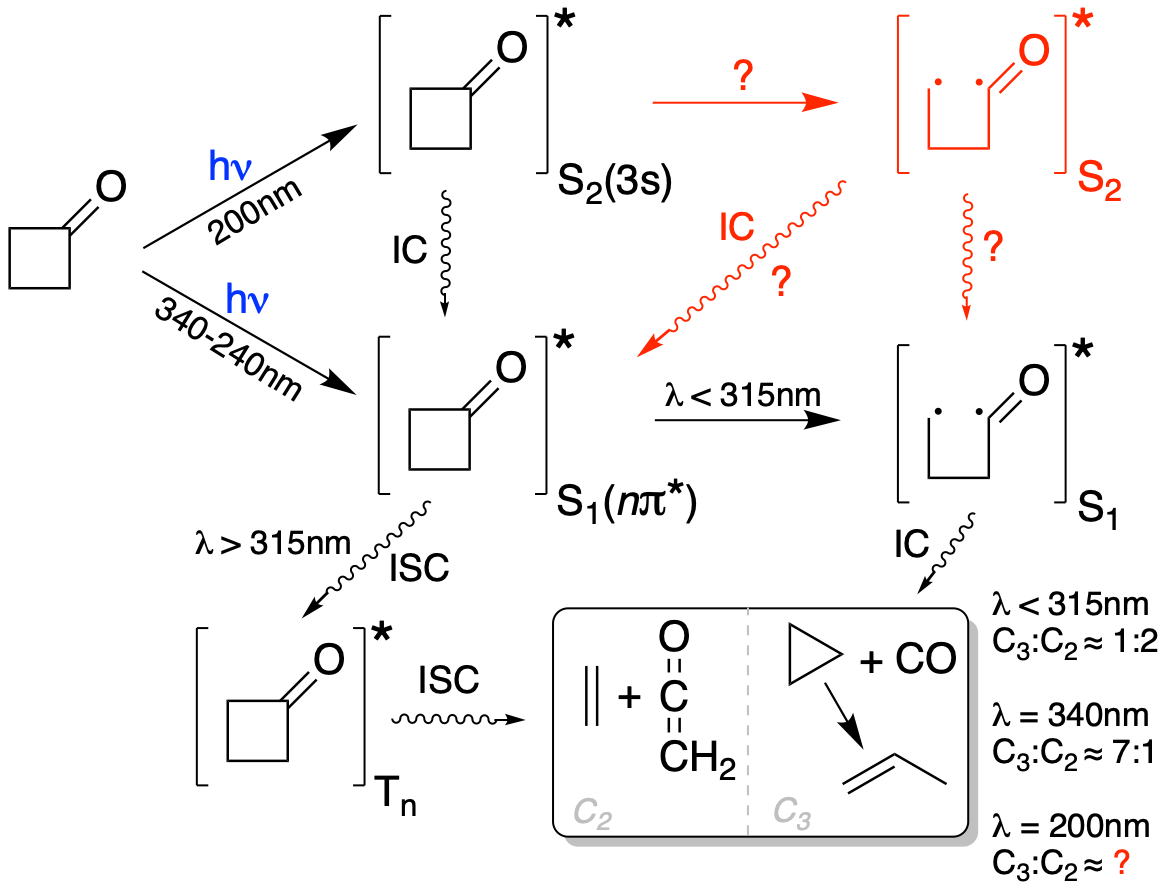}
    \caption{Schematic representation of the photochemistry of cyclobutanone triggered by photoexcitation to S$_1$($n\pi^*$) or S$_2$($3s$) as described by literature (see main text). Red symbols indicate the pathways discussed and questions addressed in the present work.}
    \label{fig:scheme}
\end{figure}

But even in the best-case scenario described above, can nonadiabatic molecular dynamics be \textit{predictive} at all? This is the central question of this Special Topic, to which this article contributes. The goal of this series of works is to predict the outcome of an ultrafast pump-probe gas-phase experiment before it is conducted. Experimentally, a molecule will be photoexcited by a short pump laser pulse at 200 nm and subsequently probed at different time delays by a probe, which in this case is a bundle of MeV electrons that will diffract on the molecule, revealing its geometrical structure. The precise details of the planned experiments are given in Sec.~\ref{sec:experiment}. The molecule chosen for this challenge is cyclobutanone, a cyclic ketone that can undergo various photochemical reactions depending on the excitation wavelength. The current knowledge of cyclobutanone photochemistry triggered by excitation in either the S$_1$ or S$_2$ electronic state is described in the following paragraphs and schematically summarized in Fig.~\ref{fig:scheme}.

The (photo)reactivity of cyclobutanone upon photoexcitation in its first excited electronic state (S$_1$) is governed by the ring strain in the four-member ring as for the other cyclic ketones.\cite{Xia2015, Kao2020} The smallest analog of the cyclic-ketone family, the highly strained cyclopropanone, undergoes an ultrafast deactivation upon excitation in its lowest excited electronic state S$_1$, releasing CO within a hundred femtoseconds.\cite{Janos2023, Cui2011} The larger and less strained analogues of this series, cyclopentanone and cyclohexanone, exhibit a slower deactivation involving intersystem crossings to the lowest triplet electronic state T$_1$  (with a $n\pi^*$ character).\cite{Xia2015, Kao2020} Cyclobutanone lies somewhere at the interface thanks to its medium ring strain, with internal conversion (IC) and intersystem crossing (ISC) processes competing when the molecule is excited in its S$_1$ electronic state. The reported photoproducts of cyclobutanone upon excitation in S$_1$ are ethene and ketene (denoted as C$_2$ products), or CO and cyclopropane/propene (coined C$_3$ products), and both families of photoproducts were observed in the gas phase and in solution.\cite{Denschlag1967, Turro1967} The ratio of photoproducts was found to vary between a photoexcitation wavelength of 340 nm (C$_3$:C$_2$ ratio 7:1) and 315 nm (C$_3$:C$_2$ ratio 1:2), but it remains constant below 315 nm.\cite{Campbell1967,Tang1976} This stark change in photoproduct quantum yields was explained by an activation barrier to ring-opening in the first excited state with a value predicted between 2 to 7~kcal/mol.\cite{Xia2015, Diau2001} When the wavelengths used produce cyclobutanone with internal energy below the barrier (340-315~nm), the direct dissociation is hampered, and ISC from S$_1$ to triplet states becomes competitive (the ISC proceeds on a nanosecond timescale\cite{Diau2001}). When photoexcited above the barrier ($<315$~nm), cyclobutanone undergoes a Norrish type-I cleavage leading to the formation of both the C$_3$ and C$_2$ products. Time-dependent experiments in solution have revealed that the dissociation happens on a subpicosecond timescale (0.65~ps),\cite{Kao2020} although previous experiments suggested a slightly longer lifetime of 5 picoseconds with 307~nm pump.\cite{Diau2001} Theoretical calculations using the \textit{ab initio} multiple spawning (AIMS) method combined with SA2-CAS(12,11)/6-31G* method predicted the S$_1$ lifetime of $\approx450$ fs and the ratio of C$_3$:C$_2$ products 1:1.\cite{Liu2016}

The photodynamics of cyclobutanone resulting from an excitation into its second excited electronic state (S$_2$, exhibiting a Rydberg 3s character in the Franck-Condon region) has been studied both experimentally\cite{Trentelman1990,Kuhlman2012a, Kuhlman2013} and theoretically.\cite{Kuhlman2012} Kuhlman and coworkers performed time-resolved mass spectroscopy (TRMS) and time-resolved photoelectron spectroscopy (TRPES) experiments on cyclobutanone, triggered by a 200~nm laser pump\cite{Kuhlman2012a} and interpreted them using multi-configuration time-dependent Hartree (MCTDH) quantum-dynamics simulations.\cite{Kuhlman2012} The time dependence of the parent ion signal from TRMS was fitted to a sequential biexponential decay with two-time constants $\tau_1=0.08\pm0.01$~ps and $\tau_2=0.74\pm0.01$~ps. The authors assumed that the longer lifetime, $\tau_2$, must reflect internal conversion to the S$_1$($n\pi^*$) state because the S$_2$ electronic state is of Rydberg character -- such an electronic-state character leads to generally bound states and a direct dissociation or isomerization of the 3s state appeared unlikely to the authors. An experimental signal corresponding to a fragment with 42 $m/z$ (cyclopropane, propylene, or ketene) yielded similar lifetimes. Based on the correspondence between the parent and fragment lifetimes, the authors stated that the processes following IC to the S$_1$($n\pi^*$) should be ultrafast. The shorter lifetime, $\tau_2$, was attributed to the formation of an intermediate structure in the S$_2$ state, probably the planarization of the initially puckered ground-state geometry. The TRPES experiments corroborated the TRMS results with similar lifetimes $\tau_1=0.31\pm0.06$~ps and $\tau_2=0.74\pm0.02$~ps. Furthermore, oscillations in the TRPES signal were recorded, exhibiting a frequency of 35~cm$^{-1}$. The MCTDH simulations supporting this experiment used a 5-dimensional vibronic coupling Hamiltonian fitted to EOM- and LR-CCSD energies.\cite{Kuhlman2012} The theoretical model includes the following modes: the carbonyl stretch to describe the changing \ce{C=O} bond length, the ring-puckering and carbonyl out-of-plane deformation to describe the changing structure of the ring, and the angle between the \ce{C=O} bond and the \ce{C-C-C} plane combined with ring modes affecting the length of the \ce{C-C} bonds. The calculated decay of the S$_2$-photoexcited cyclobutanone to the S$_1$ electronic state was also fitted by a biexponential function, leading to time constants of 0.95~ps and 6.32~ps. The lifetimes observed for the IC process were attributed to (\textit{i}) an immediate activation of coupling modes leading to a fast population transfer from S$_2$ towards S$_1$ (direct mechanism), and  (\textit{ii}) an internal vibrational energy redistribution followed by a slower population transfer to S$_1$ (indirect mechanism). We note that no population reached the ground electronic state S$_0$ or the S$_3$(3p) electronic state in this dynamics. The trapping in the S$_1$ state is not surprising as these simulations are restricted by the underlying harmonic approximation of the linear vibronic coupling model and cannot describe direct bond dissociation. This restriction also hampered a possible dissociation in the S$_2$.
Another experiment by Trentelman and coworkers\cite{Trentelman1990} measured rotational spectra of \ce{CO} fragments following photoexcitation of cyclobutanone at 193~nm, revealing two groups of fragments with different rotational temperatures. The high-energy fragments ($\approx 3000$~K) were attributed to direct dissociation into CO and cyclopropane. The low-temperature fragments ($\approx 200$~K) were attributed to the formation of an excited ketene, which then dissociates into \ce{CH2} and CO. The ratio of these channels was found to be 85:15. The authors estimated the ratio of C$_3$:C$_2$ products to be 57:43 based on the assumption that cyclobutanone in S$_2$ undergoes an ultrafast IC to the S$_1$ state, which then drives the dissociation, and extrapolation of the quantum yields in the S$_1$ state. 

Moving to the current prediction challenge, the time-resolved MeV-UED experiment on cyclobutanone \textit{in vacuo} will use a 200 nm pump pulse to promote the molecule in its S$_2$($3s$) state. Time-resolved MeV-UED experiments are sensitive to the geometry of the molecule probed at a given time following photoexcitation\cite{yang2018imaging,wolf2019photochemical,doi:10.1146/annurev-physchem-082720-010539} (electronic information can be obtained too in some cases\cite{doi:10.1126/science.abb2235}) and therefore offers a perfect tool to challenge the nuclear wave packets produced by nonadiabatic molecular dynamics simulations and unravel the actual mechanistic details of the photodissociation that have eluded earlier experiments (question marks in Fig.~\ref{fig:scheme}). In the following, we propose an \textit{in silico} version of this experiment, where we simulate the excited-state dynamics of cyclobutanone and the subsequent dynamics of its photoproducts in the ground electronic state from the perspective of nonadiabatic molecular dynamics. We also predict the MeV-UED signal stemming from our simulations. The time constraints imposed by this prediction challenge mean that we had to opt for a compromise between computational efficiency and accuracy -- XMS-CASPT2 for the electronic structure and trajectory surface hopping (TSH) for the nonadiabatic molecular dynamics -- and that the number of tests we would usually conduct to validate the electronic-structure methods had to be kept to a minimum. We therefore supplement our results section with a discussion on the potential limitations of our model (Sec.~\ref{sec:limitations}) and their possible influence on the results of our simulations.  

\section{Methods}
\label{methodsection}

\subsection{Electronic-structure methods and benchmark}

The electronic structure of cyclobutanone in this work was mainly described by the extended multistate complete active space second-order perturbation theory (XMS-CASPT2)\cite{Park2017} based on a state-averaged complete active space self-consistent field (SA-CASSCF) reference wavefunction as implemented in the BAGEL package.\cite{Shiozaki2018} The state-averaging procedure considered three singlet states with equal weights. The active space comprised 8 electrons in 8 orbitals, namely two $\sigma_\mathrm{CC}$ adjacent to carbonyl, $\pi_\mathrm{CO}$, $n_\mathrm{O}$, $\pi_\mathrm{CO}^*$, $Ry(3s)$, and two corresponding $\sigma^*_\mathrm{CC}$ orbitals (see Fig.~S1 for a representation of the active-space orbitals). All calculations used the aug-cc-pVDZ basis set combined with cc-pVTZ-jkfit auxiliary basis for the density fitting. This overall level of electronic-structure theory is denoted XMS-CASPT2(8/8) in the following and was used to optimize the electronic-state minima, transition states (TSs), and minimum energy conical intersections (CIs) of cyclobutanone, see the supplementary material (SM) for the corresponding Cartesian coordinates. Linear interpolations in internal coordinates (LIICs) between these critical geometries were used to investigate the potential energy surfaces and further benchmark the level of electronic-structure theory. In particular, the XMS-CASPT2(8/8) strategy described above was thoroughly compared (excitation energies and LIICs) to equation-of-motion coupled-cluster singles doubles (EOM-CCSD) and linear-response time-dependent density functional theory (LR-TDDFT) employing the Tamm-Dancoff approximation (TDA) with the \textomega B97XD exchange-correlation functional. Both the EOM-CCSD and LR-TDDFT/TDA/\textomega B97XD methods were performed in Gaussian 09, Revision D.01.\cite{g09} The excitation energies were also benchmarked against the EOM-CC3 method\cite{Paul2021} implemented in the eT code.\cite{Folkestad2020} We note that the presence of a carbonyl group in the studied molecule and the importance of a low-lying electronic state with a $n\pi*$ character hampers the use of ADC(2).\cite{D1CP02185K} See the SM for additional details of this comparison not discussed in the main text.

Møller–Plesset second-order perturbation theory (MP2) as implemented in Gaussian 09, Revision D.01\cite{g09} was also used to model the ground state and sample ground-state nuclear density of cyclobutanone and its photoproducts.  For optimization of the photoproducts (CO, cyclopropane, propene, ketene, and ethene) and also ground-state continuation of the nonadiabatic molecular dynamics simulations (see below), we used the MP2/aug-cc-pVDZ level of theory. This strategy was benchmarked against XMS-CASPT2 and unrestricted PBE0 (UPBE0) calculations (see SM). The UPBE0 calculations were done in Gaussian 09, Revision D.01.\cite{g09} with a mixed guess for the initial wave function. For sampling the ground-state nuclear density with molecular dynamics, we employed the MP2 method with a smaller basis set cc-pVDZ.

Spin-orbit coupling (SOC) interactions were modeled at the SA-CASSCF(8/8) level of theory, state-averaging over three singlet and three triplet states with equal weights [SA(3S,3T)-CASSCF(8/8)] using the same active space as described above and the Molpro2012 package.\cite{molpro2012} The energies of the triplet states (and their character) obtained at this level of theory were benchmarked against the EOM-CCSD and LR-TDDFT/TDA/\textomega B97XD methods (see SM). The SOC matrix elements were evaluated using the Breit--Pauli Hamiltonian $\hat{H}_\mathrm{SOC}$ and the values reported here correspond to the magnitude of the SOC between a given singlet electronic state S$_n$ and the M$_S$ sublevels of a triplet electronic state T$_m$:

\begin{equation*}
    H_\mathrm{SOC}^\mathrm{T_m,S_n} = \sqrt{\sum_{\mathrm{M_S}=-1}^1 \left|\left\langle \psi_\mathrm{T_m,M_S} \middle| \hat{H}_\mathrm{SOC}  \middle| \psi_\mathrm{S_n,0} \right\rangle \right|^2}
\end{equation*}

\subsection{Photoabsorption cross-section}

Two approaches were used to simulate the photoabsorption cross-section of cyclobutanone: (\textit{i}) Franck-Condon Herzberg-Teller (FCHT) method to obtain an accurate depiction of the vibronic structure of the $\mathrm{S}_2\leftarrow\mathrm{S}_0$ band and (\textit{ii}) the nuclear ensemble approach (NEA) to reproduce the shape of this band and sample initial conditions for the nonadiabatic molecular dynamics simulations.\cite{Crespo-Otero2012}

We calculated the vibrationally-resolved absorption spectrum of cyclobutanone using the FCHT terms at the LR-TDDFT/TDA/\textomega B97XD/aug-cc-pVDZ level of theory, as implemented in Gaussian 09, Revision D.01.\cite{g09} This calculation employed a harmonic approximation around the optimized minima of the ground electronic state S$_0$ and the second excited electronic state S$_2$, and considered a temperature of 298~K. We note that the benchmark described above and presented in the SM indicates that LR-TDDFT/TDA/\textomega B97XD/aug-cc-pVDZ offers an adequate description of S$_2$ minimum. 

The NEA was used to capture the envelope of the $\mathrm{S}_2\leftarrow\mathrm{S}_0$ band of the photoabsorption cross-section.\cite{Crespo-Otero2012} The ground-state nuclear probability density of cyclobutanone was represented by 5000 geometries that were sampled from \textit{ab initio} Born--Oppenheimer molecular dynamics (BOMD) at the MP2/cc-pVDZ with our molecular dynamics code ABIN.\cite{Hollas2019} Both thermal effects (298~K) and zero-point energy effects were included by the quantum thermostat (QT) based on the generalized Langevin equation.\cite{Ceriotti2009,Ceriotti2010} The cyclobutanone molecule was first equilibrated and then propagated for 96~ps with a time step of 20~a.u. ($\sim$ 0.5~fs). Recent studies suggest that QT-BOMD offers a more reliable strategy to approximate the ground-state nuclear probability of flexible molecules.\cite{prlj2021calculating,doi:10.1021/acs.jpca.3c02333} The photoabsorption cross-section spectra were obtained using LR-TDDFT/TDA/\textomega B97XD/aug-cc-pVDZ excitation energies ($\Delta E_{\mathrm{S}_0\leftarrow\mathrm{S}_2}$) and transition dipole moments ($\boldsymbol{\mu}_{\mathrm{S}_0\leftarrow\mathrm{S}_2}$) between S$_0$ and S$_2$ on the support of the 5000 geometries extracted from each distribution, using

\begin{equation*}
    \sigma_{\mathrm{S}_2\leftarrow\mathrm{S}_0}(E)=\frac{\pi}{3 \hbar \varepsilon_0 c}\frac{1}{N_s h \sqrt{2\pi}} \sum_{i=1}^{N_s}  \Delta E_{\mathrm{S}_0\leftarrow\mathrm{S}_2,i} |\boldsymbol{\mu}_{\mathrm{S}_0\leftarrow\mathrm{S}_2,i}|^2\exp{\left(-\frac{\left(E-\Delta E_{\mathrm{S}_0\leftarrow\mathrm{S}_2,i}\right)^2}{2h^2}\right)},
\end{equation*}

\noindent where $N_s$ is the number of sampled geometries (5000 in the present case) and $h$ is the the Gaussian broadening factor, which was set according to Silverman's rule of thumb

\begin{equation*}
    h\approx \left( \frac{4s^5}{3N_s}\right)^{\frac{1}{5}}
\end{equation*}

\noindent where $s$ is the standard deviation of the excitation energies.\cite{Srsen2018, Silverman1986}

\subsection{Nonadiabatic molecular dynamics}

The photodynamics of cyclobutanone upon photoexcitation at 200 nm was simulated with two versions of the trajectory surface hopping (TSH) method implemented in the ABIN code\cite{Hollas2019}: the fewest-switches surface hopping (FSSH)\cite{Tully1990} and Landau-Zener surface hopping (LZSH).\cite{Suchan2020} The initial conditions (nuclear positions and momenta) were selected from the 5000 geometries sampled from the MP2/cc-pVDZ QT-BOMD simulations described above. An energy filter\cite{Martinez-Mesa2015} was applied to the 5000 sampled geometries: only the geometries with a S$_2$$\leftarrow$S$_0$ electronic transition falling within an energy window between 6.1764 and 6.2220~eV -- 200.74 and 199.27~nm respectively -- were selected. The applied energy window reflects the energy width of the pulse that will be used in the experiment (see below). The 119 initial conditions selected by the energy windowing were then vertically promoted to the S$_2$ electronic state, where the trajectories were initiated. For the FSSH dynamics, the energy-based decoherence scheme was applied using the recommended 0.1~a.u. value.\cite{Granucci2007} The nonadiabatic transitions were mediated by the nonadiabatic coupling vectors (the latter being calculated only when the energy difference between coupled electronic states dropped below 1.5~eV). The nuclear velocities were rescaled along the nonadiabatic coupling vector after a successful hop, and the time step was set to 10 a.u. ($\sim$0.25 fs). The total energy conservation was carefully monitored for all trajectories. The active state chosen for the XMS-CASPT2(8/8) method was stable until the molecule opened its ring, cleaving the \ce{C-C} bond adjacent to the carbonyl group. When this process happened, the Rydberg $3s$ state was pushed higher in energy, forcing the Rydberg orbital to rotate out of the active space slowly. Nevertheless, the orbital rotations causing jumps in total energy occurred mainly when the molecule was already evolving in the ground electronic state. This means that total energy was well conserved in the segment of the dynamics evolving in S$_2$, but suffered moderate jumps of tenths of eV in total energy when the trajectories were evolving in S$_0$ and, in some cases, in S$_1$. We note that a larger active space, XMS(4)-CASPT2(8/9), would be more robust and stable but computationally intractable (in particular, given the time constraints of this prediction challenge).

To collect 2 ps of simulation time for all trajectories, we switched from the time-consuming FSSH/XMS-CASPT2 level of theory to \textit{ab initio} BOMD with MP2/aug-cc-pVDZ (time step of 10~a.u) once the molecule reached the ground electronic state and left the nonadiabatic region. The criteria for switching from FSSH/XMS-CASPT2 to BOMD/MP2 were \textit{i}) a formation of stable C$_2$ or C$_3$ products and \textit{ii}) a distance between the nearest carbon atoms of the two dissociated fragments got bigger than 8~{\AA}. When a ground-state cyclobutanone was formed, the FSSH/XMS-CASPT2 simulations were continued. We coin this strategy of combining nonadiabatic and Born--Oppenheimer molecular dynamics (NA+BO)MD in the following.
We note that the strategy of switching between nonadiabatic molecular dynamics to BOMD was successfully used (and benchmarked) in earlier studies.\cite{mignolet2016rich,pathak2020tracking} In 8 cases, SA-CASSCF failed to converge before the dissociated products departed far enough to fulfil criterion \textit{ii}). In these cases, we switched to a BOMD at the geometry where the SA-CASSCF convergence issue was experienced. We stress that, for all trajectories, stable ground-state C$_2$ or C$_3$ products were always formed by the time the switch to BOMD was operated.

\subsection{Analysis of the nonadiabatic molecular dynamics simulations}

The time-dependent electronic-state populations were evaluated as an average of the FSSH active electronic states over all trajectories, i.e. the states on which the molecule evolves. We estimated the error bars of the $i$-th electronic-state population at a given time $t$ using a multinomial distribution as $\epsilon_i(t)=z\sqrt{\frac{1}{N}p_i(t)\left(1-p_i(t)\right)}$, where $N$ is the number of trajectories, $p_i(t)$ is the population of $i$-th electronic state at time $t$, and $z=1.96$ corresponds to a 95 \% confidence interval. To account for the temporal smearing of the results, the calculated populations $p_i(t)$ were then convoluted with a normalized Gaussian envelope $\chi_\mathrm{pul.env.}$ of the excitation pulse (described in the experimental details) based on 

\begin{equation*}
    \Tilde{p}_i(t) = \int_{-\infty}^\infty {p}_i(t-\tau) \chi_\mathrm{pul.env.}(\tau) \mathrm{d}\tau
\end{equation*}

\noindent To fit the population traces and obtain the lifetimes, we considered a consecutive reaction scheme $\mathrm{S}_2 \rightarrow \mathrm{S}_1 \rightarrow \mathrm{S}_0$. For more details on the fitting and unconvoluted population traces, see SM.

\begin{figure}[ht]
    \centering
    \includegraphics[width=0.5\textwidth]{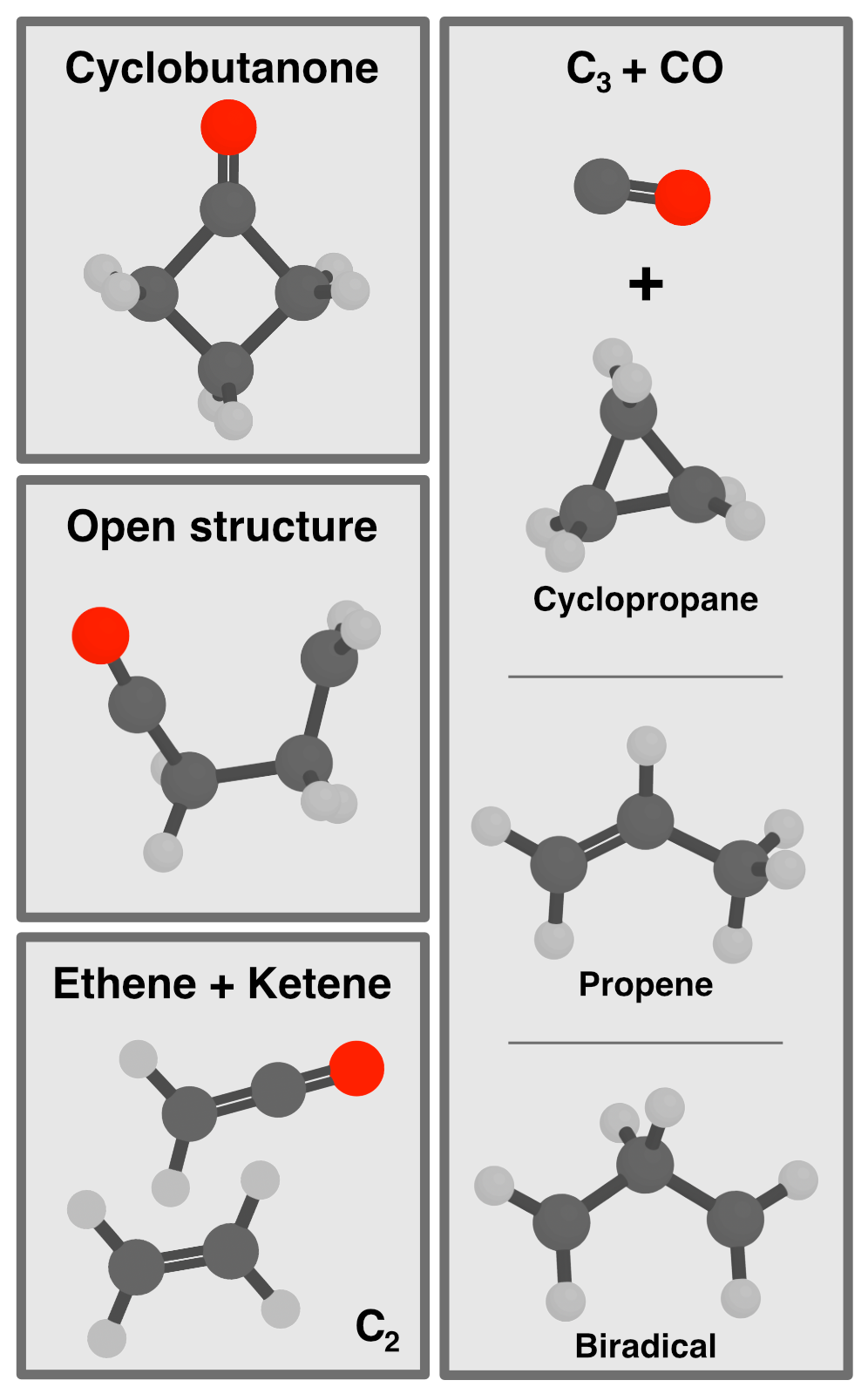}
    \caption{The different categories of (photo)products considered in our analysis of nonadiabatic dynamics.}
    \label{fig:structures}
\end{figure}

To calculate the time-dependent relative populations of the different photoproducts, we developed a classification scheme that sorts molecular geometries extracted from the trajectories into six possible photoproduct categories (see Fig.~\ref{fig:structures}). We distinguish the following (photo)products: \ce{CO} + cyclopropane, \ce{CO} + propene, \ce{CO} + \ce{^{.}CH2-CH2-CH2^{.}} biradical, ethene + ketene, any transient/intermediate opened structures, and cyclobutanone, which is both initial structure and one of the photoproducts when ring-closing happens. The classification scheme used a series of geometrical parameters (see Fig.~S5 in the SM for the parameters and categories) and was validated by visual analysis of the trajectories. Its sensitivity to the choice of parameters was tested and this scheme could capture all the types of structures encountered during the nonadiabatic and ground-state dynamics. 
The quantum yield for a given (photo)product $k$, $\phi_\mathrm{k}(t)$, was evaluated as $\phi_\mathrm{k}(t)=\frac{N_k(t)}{N}$, where $N$ is the total number of trajectories and $N_k(t)$ the number of trajectories showing the (photo)product $k$ at time $t$. The statistical error was derived from a multinomial distribution as $\epsilon_k=z\sqrt{\frac{1}{N}\phi_\mathrm{k}(t)\left(1-\phi_\mathrm{k}(t)\right)}$, where $z=1.96$ is used to express a 95\% confidence interval. The resulting populations were again convoluted with the experimental pulse to account for the temporal smearing of the pulse.

Finally, we note that we tested weighting the result by transition dipole moments, but we did not observe any noticeable effects on the resulting quantities.

\subsection{Details of the planned experiment}
\label{sec:experiment}

We summarize here the main details of the time-resolved MeV-UED experiment that will be conducted on cyclobutanone. A complete description of the prediction challenge is available online.\cite{predictionchallenge} The gas-phase sample of cyclobutanone will be irradiated with 200 nm light ($\approx80$ fs cross-correlation), and electron-diffraction images will be obtained with an expected 150-fs time resolution (FWHM) and 0.6 Å spatial resolution (2$\pi$/S\textsubscript{max}). The scattering vector S will range from 1-10 Å$^{-1}$. Diffraction images will be collected for time delays ranging from -1 ps to tens of ps in variable step sizes. The region around time zero (-200 fs to 200 fs) will be scanned with 30-fs step sizes. Longer positive delays will be scanned with step sizes of up to several ps. The intensity of the 200 nm excitation light will be kept as low as possible (5 μJ) to avoid multiphoton excitation, exciting $\sim10\%$ of the molecules. 

\subsection{Calculation of UED signal}
The theoretical static pair-distribution function [$\text{PDF}(r)$] of all contributing interatomic distances $r$ and time-dependent fractional change signal ($\Delta I/I(s,t)$, with $I(s)$ the one-dimensional scattering intensity and $s$ the momentum transfer vector) and difference pair distribution functions [$\Delta\text{PDF}(r,t)$] signals reported in this work were calculated within the independent atom model\cite{iam1,iam2} (IAM) using the nuclear geometries from LIICs and nuclear geometries obtained along the entire swarm of FSSH trajectories (see Ref.~\citenum{nunes2023monitoring} and its SM for the detailed equations). 

\subsection{Carbon footprint of the calculations presented in this work}

The simulations in this work used approximately 145'920 CPU hours (AMD EPYC 7513), drawing $\sim$2.26 MWh. Based in the Czech Republic, this consumption leads to a carbon footprint of 1.12 T CO\textsubscript{2}e, which is equivalent to a carbon sequestration of 101.86 tree-years or the emission of a return flight between New York City and San Francisco (calculated using green-algorithms.org v2.2\cite{carbonfootprint}).

\section{Results and discussion}

We start this Section by discussing a static picture of the photodynamics of cyclobutanone (Sec.~\ref{staticpart}), where we first confirm the nature of the excited electronic state reached with the experimental 200 nm laser pulse. Then, we introduce the possible photochemical pathways of the molecule in terms of LIICs and discuss the mechanism. Finally, we present a steady-state picture of the expected MeV-UED experimental signal produced along the two proposed pathways. In the second part of this Section (Sec.~\ref{timeresolved}), we move to a time-resolved picture and discuss the results of our FSSH simulations before calculating the time-resolved signal that our nonadiabatic molecular dynamics trajectories predict for the first 2 ps of dynamics following photoexcitation of cyclobutanone at 200 nm.

\subsection{Static picture of the photochemistry of cyclobutanone}
\label{staticpart}
\subsubsection{Photoabsorption cross-section of cyclobutanone}
\label{sec:photoabs}
\begin{figure}[ht]
    \centering
    \includegraphics[width=0.8\textwidth]{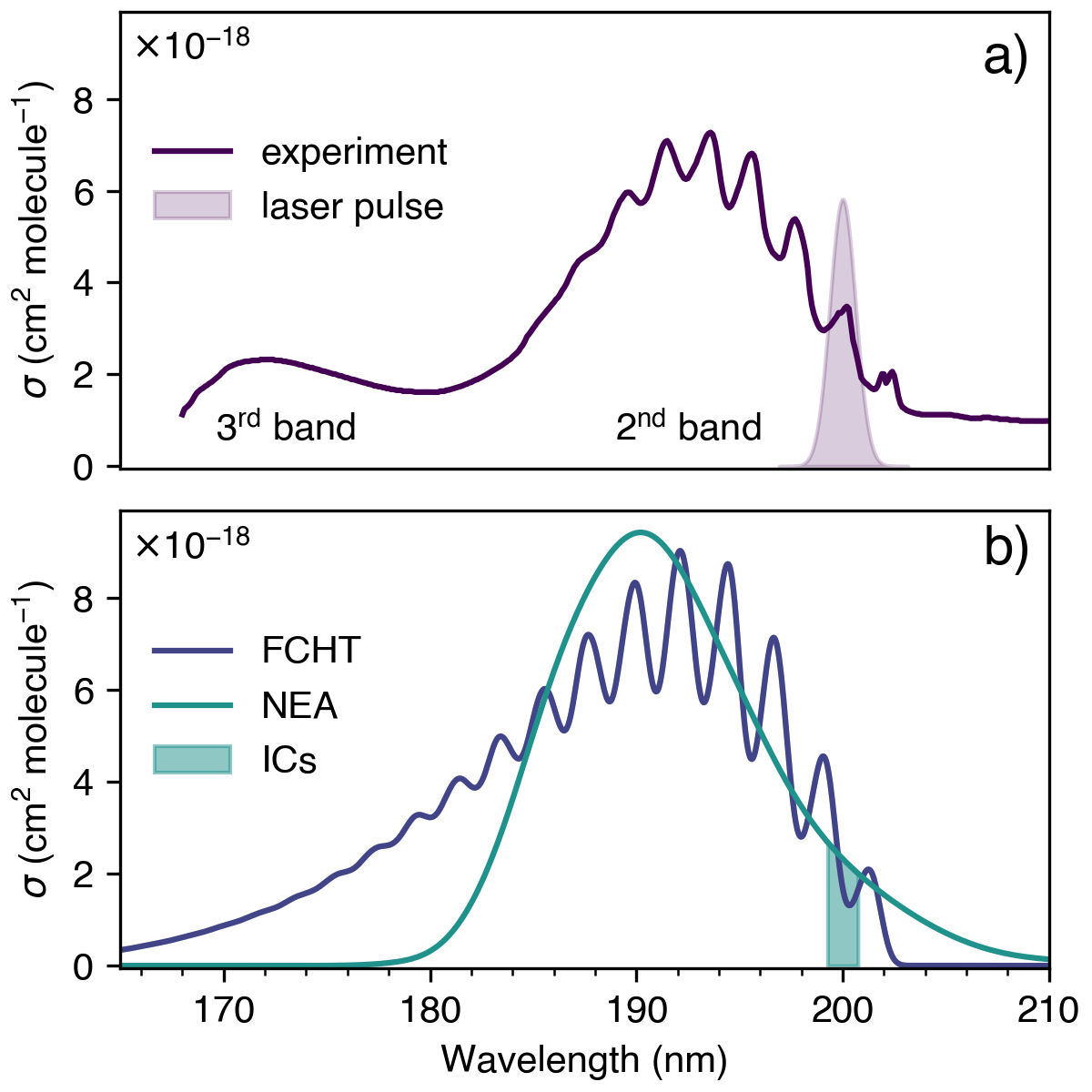}
    \caption{Experimental and theoretical photoabsorption cross-sections of cyclobutanone. a) Experimental photoabsorption cross-section measured by Udvarhazi and El-Sayed\cite{Udvarhazi1965} and digitized by the MPI-Mainz UV/VIS Spectral Atlas of Gaseous Molecules of Atmospheric Interest.\cite{Keller-Rudek2013} The energy distribution of the experimental pump laser pulse (200 nm, $\approx80$ fs cross-correlation) is superimposed to highlight the region of photoexcitation (see Sec.~\ref{methodsection}). b) Calculated S$_2$$\leftarrow$S$_0$ photoabsorption cross-section for cyclobutanone using the FCHT and the NEA strategies. In both cases, excitation energies and all required electronic quantities were calculated with LR-TDDFT/TDA/\textomega B97XD. For the NEA, 5000 geometries were sampled from QT-BOMD. The highlighted area (ICs) represents the energy window employed to select the initial conditions for the subsequent nonadiabatic molecular dynamics, mimicking the role of the experimental laser pulse. No shifts were applied to the calculated photoabsorption cross-sections.}
    \label{fig:spec}
\end{figure}

The experimental 200~nm pump laser pulse targets the low-energy region of the second absorption band in the absorption spectrum of cyclobutanone (Fig.~\ref{fig:spec}a). This absorption band is well separated from the first one, which is located at around 280~nm, but it slightly overlaps with the third band visible in the spectrum near 175 nm. 

Let us begin by assigning the character of the electronic states responsible for these bands and validate the selected electronic structure methods. We calculated the first five excitation energies at the optimized geometry of the ground-electronic state (MP2/cc-pVTZ) with various electronic-structure methods (Table~\ref{tab:excenergies}). The excitation to the first excited electronic state S$_1$, producing the low-energy absorption band, corresponds to a $n\pi^*$ transition located on the carbonyl moiety of cyclobutanone, with a corresponding very low oscillator strength. The second absorption band belongs solely to a transition towards the S$_2$ electronic state, which exhibits a $nRy(3s)$ character, i.e. transition from the non-bonding orbital to the Ryberg 3s orbital. The electronic-state character will be denoted $3s$ in the following. The electronic energy of the S$_2$ electronic state is well separated from that of the other states in the Franck-Condon region and appears to be the target of our calculations -- we will confirm this assumption below by calculating the photoabsorption cross-section corresponding to this transition. The following peak located at 175~nm is attributed to three nearly-degenerate transitions to the excited electronic states S$_3$-S$_5$, all three transitions showing a $nRy(3p)$ character. Comparing the transition energies and oscillator strengths obtained with different electronic-structure methods (Table~\ref{tab:excenergies}), we observe a consistent performance between the coupled-cluster-based techniques, XMS-CASPT2(8/8) and LR-TDDFT/TDA -- XMS-CASPT2 underestimating the oscillator strength for the brighter transitions, as observed for other molecular systems.\cite{doi:10.1021/acs.jpca.3c02333} Given our interest in further validating the nature of the second absorption band by reproducing its vibronic structure and obtaining consistent initial conditions with the NEA method, we chose to use LR-TDDFT/TDA/\textomega B97XD for the calculation of photoabsorption cross-section calculations that will be reported later in this Section. The SM contains a more extensive benchmark of the electronic-structure methods, including also the role of triplet electronic states and a larger basis set, and performance in the vicinity of the S$_2(3s)$ minimum-energy geometry.

\begingroup
\squeezetable
\begin{table}[ht]
\begin{tabular}{m{1.0cm} m{1.0cm} m{0.72cm} m{0.88cm} m{0.75cm} m{0.75cm}m{0.75cm}m{0.75cm}m{0.75cm}m{0.75cm}}
\hline
\multirow{2}{*}{Transition} & \multirow{2}{*}{Character} & \multicolumn{2}{c}{EOM-CC3}  & \multicolumn{2}{c}{EOM-CCSD}  & \multicolumn{2}{c}{\makecell{LR-TDDFT/TDA\\\textomega B97XD}} & \multicolumn{2}{c}{XMS-CASPT2(8/8)} \\
 & & $\Delta E$ (eV) & \textit{$f$} & $\Delta E$ (eV) & \textit{$f$} & $\Delta E$ (eV)   & \textit{$f$} & $\Delta E$ (eV)   & \textit{$f$} \\
\hline
S$_1$$\leftarrow$S$_0$ & $n\pi^*$ & 4.39 &$4\cdot10^{-7}$& 4.41 & 0.0000 & 4.38 & 0.0000 & 4.37 & 0.0000 \\
S$_2$$\leftarrow$S$_0$ & $nRy(3s)$ & 6.36 & 0.0380 & 6.43 & 0.0378 & 6.58 & 0.0376 & 6.25 & 0.0167 \\
S$_3$$\leftarrow$S$_0$ & $nRy(3p)$ & 6.96 & 0.0016 & 7.05 & 0.0022 & 7.15 & 0.0019 & - & - \\
S$_4$$\leftarrow$S$_0$ & $nRy(3p')$ & 7.12 & 0.0000 & 7.19 & 0.0001 & 7.33 & 0.0001 & - & - \\
S$_5$$\leftarrow$S$_0$ & $nRy(3p'')$ & 7.15 & 0.0021 & 7.23 & 0.0016 & 7.33 & 0.0003 & - & - \\
\hline
\end{tabular}
\caption{Excitation energies ($\Delta E$) and oscillator strengths ($f$) obtained at different levels of electronic-structure theory on the support of the ground-state geometry of cyclobutanone (MP2/cc-pVTZ). All excited-state methods employed the aug-cc-pVDZ basis set.}
\label{tab:excenergies}
\end{table}
\endgroup

Before discussing the photoabsorption cross-section of cyclobutanone, it is crucial to briefly mention its ground-state geometry. Cyclobutanone is not planar in its ground-electronic state and, as mentioned in the Introduction, the far-infrared and microwave spectra of cyclobutanone revealed a low vibrational mode with a frequency of 35 cm$^{-1}$ that corresponds to a ring-puckering motion.\cite{Borgers1966,Scharpen1968,Alonso1992} A fit of the spectral lines indicates a double-well ground-state potential for the inversion of the ring with a barrier of 4.8-7.6~cm$^{-1}$. The zero-point energy of the double-well potential is approximately 9.2~cm$^{-1}$ above the barrier.\cite{Scharpen1968} This observation is particularly important for the sampling of the ground-state nuclear density using a harmonic approximation. We tested different electronic-structure methods and found a wide range of S$_0$-minimum geometries and transition-state energies, indicating that the ring-puckering appears to be highly sensitive to the description of dispersion interactions offered by a given method (see SM for a summary of these tests). Although small, these effects strongly influence the shape of the ground-state potential energy surface and, as such, the sampling of geometries representative of the ground-state nuclear probability density. Our benchmark shows that MP2 was in agreement with CCSD and XMS-CASPT2, as was DFT with the \textomega B97XD functional, which gives the best result among the tested exchange-correlation functionals. We note that the second lowest vibrational mode with a frequency of 395~cm$^{-1}$ appears to be harmonic.

We now move to the simulation of the photoabsorption cross-section for the second band of cyclobutanone. To reproduce the experimentally-observed vibronic structure of this band (Fig.~\ref{fig:spec}a), we first used the FCHT method based on DFT/\textomega B97XD (ground-state optimized geometry) and LR-TDDFT/TDA/\textomega B97XD (the minimum of the S$_2$ electronic state, transition energy, transition dipole moment and its nuclear derivative). While the S$_0$ ring-puckering mode is not adequately described within the harmonic approximation, the target S$_2$ electronic state is harmonic. Therefore, the vibrational splitting in the photoabsorption cross-section should not be affected as it depends only on the vibrational energy levels of the S$_2$ electronic state. The ground-state anharmonicity of the ring puckering mode may, however, impact the absolute shift of the photoabsorption cross-section and its intensity. The FCHT photoabsorption cross-section is presented in Fig.~\ref{fig:spec}b and, despite the potential anharmonicity issue, shows an excellent agreement with the experimental cross-section, capturing the first few peaks of the vibronic progression with accuracy. This result further confirms the electronic character of this band and validates the electronic-structure methods employed in this work. We stress that no shifts were applied to the calculated photoabsorption cross-sections presented here. The modes responsible for the peak in the photoabsorption cross-section at around 200~nm -- excited by the pump pulse -- correspond to a ring puckering and \ce{C=O} stretching.

We also calculated the photoabsorption cross-section using the NEA, which proposes to use the reflection principle based on a set of sampled geometries representative of the ground-state nuclear probability density. The NEA is faster to deploy than the FCHT but does not give access to the vibronic structure, as the nuclear wavefunctions for the excited electronic state of interest are not calculated. The sampling of nuclear coordinates required by the NEA means that this technique also offers a simple strategy to select initial conditions for any subsequent nonadiabatic molecular dynamics (nuclear momenta are then also required). The ground-state probability density was sampled using QT-BOMD (see Sec.~\ref{methodsection}), which is able to describe anharmonic potential energy surfaces such as the ring puckering mode of cyclobutanone. QT-BOMD appears to outperform the other common sampling strategies like the harmonic Wigner distribution for flexible molecules.\cite{doi:10.1021/acs.jpca.3c02333} The QT-BOMD simulation used MP2/cc-pVDZ for the ground-state electronic-structure quantities, while the excitation energy and transition dipole moment required for the NEA was obtained with LR-TDDFT/TDA/\textomega B97XD for each sampled geometry. The resulting NEA photoabsorption cross-section was constructed from 5000 geometries and recovers the low-energy spectral envelope of the FCHT photoabsorption cross-section (Fig.~\ref{fig:spec}b). To select the initial conditions (nuclear positions and momenta)  for the nonadiabatic molecular dynamics simulations, we used an energy window (box in Fig.~\ref{fig:spec}b) mimicking the excitation laser pulse to filter the geometries in the photoexcitation region of interest.

\subsubsection{Possible relaxation pathways of cyclobutanone upon photoexcitation at 200~nm}

\begin{figure}[ht]
    \centering
    \includegraphics[width=1.0\textwidth]{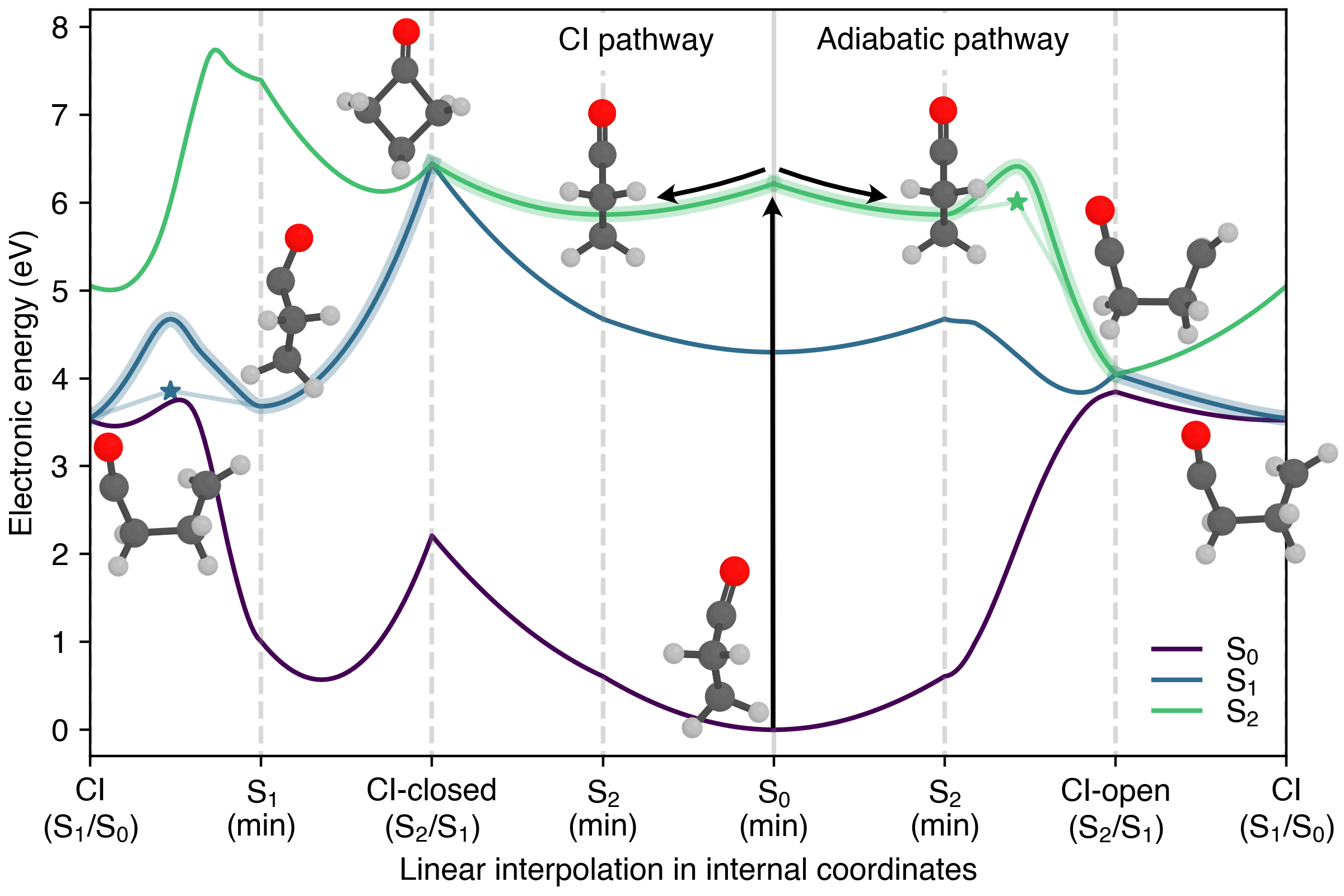}
    \caption{Two possible relaxation pathways upon photoexcitation of cyclobutanone in S$_2$ (vertical arrow) from its S$_0$ minimum. Electronic energies (XMS-CASPT2(8/8)/aug-cc-pVDZ) are represented along the LIICs characterizing two relaxation pathways -- the CI pathway [left of S$_0$(min)] and the adiabatic pathway [right of S$_0$(min)]. As LIICs do not accurately describe barrier energies, we denote the energies of the true transition state with a star. A shaded area highlights the electronic state likely to drive the dynamics of the molecule from S$_2$ to S$_0$. The molecular structure representative of each located critical point is superimposed on the pathways (the spatial orientation of each structure was chosen to highlight the specific distortions of the molecular framework).}
    \label{fig:liic}
\end{figure}

Upon photoexcitation with a 200~nm laser pulse, our earlier results indicate that cyclobutanone will be promoted into its S$_2(3s)$ electronic state. After an initial relaxation away from the Franck-Condon region and towards the planar S$_2$ minimum, two different relaxation pathways appear to be possible to reach S$_1$ and S$_0$ (these pathways will be further confirmed by our nonadiabatic molecular dynamics simulations presented below). To map the energy profile along these pathways, we optimized the critical geometries characterizing each pathway -- energy minima (min) and conical intersections (CIs) – at the XMS-CASPT2(8/8)/aug-cc-pVDZ and connected these geometries by interpolated structures using LIICs. We then used XMS-CASPT2(8/8) to obtain the electronic energies for S$_0$, S$_1$, and S$_2$ on the support of the LIICs geometries -- see Fig.~\ref{fig:liic} for the final result (the electronic-structure benchmark validating the use of XMS-CASPT2(8/8) can be found in the SM). The first pathway -- coined the 'CI pathway' (left part of Fig.~\ref{fig:liic}) -- shows how cyclobutanone can funnel from S$_2$(min) to S$_1$ preserving its closed ring via the CI-closed(S$_2$/S$_1$) conical intersection (0.58~eV above S$_2$(min)), after which the molecule can relax towards the S$_1$ minimum. From the S$_1$ minimum, cyclobutanone can open the ring, overcoming a barrier of approximately 0.18 eV in S$_1$, and hit the CI(S$_1$/S$_0$) conical intersection, which takes the molecule back to the ground electronic state, leading to the formation of photoproducts or possibly the reformation of cyclobutanone via ring closure. 

The second nonradiative pathway involves a change of electronic character along the S$_2$ electronic state, leading to the direct ring opening of cyclobutanone after overcoming a barrier of about 0.15~eV -- this pathway is coined the 'adiabatic pathway' (right part of Fig.~\ref{fig:liic}). We note that the S$_3$ state stays energetically separated from S$_2$ in the proximity of the S$_2$ transition state, limiting the possibilities for a nonadiabatic transition towards S$_3$ (see SM). Since the activation barrier for this process is smaller than the activation barrier for the CI pathway, one may expect that this pathway dominates the nonradiative decay of cyclobutanone. Once the ring-opened molecule crosses the S$_2$ transition state, it decays toward the CI-open(S$_2$/S$_1$) and can then funnel through CI(S$_1$/S$_0$) (the same conical intersection as for the CI pathway). Interestingly, the ring-opening process in S$_2$ means that the three electronic states considered become close in energy in the direct vicinity of CI-open(S$_2$/S$_1$) (see electronic energies near CI-open(S$_2$/S$_1$) in Fig.~\ref{fig:liic}). This proximity between S$_2$, S$_1$, and S$_0$ may lead to efficient transfers through the three states in this region of configuration space. We will see that this prediction is corroborated by the nonadiabatic molecular dynamics simulations described below. 

At this stage, we should comment on the possibility of ISC processes in the photodynamics of cyclobutanone following photoexcitation at 200~nm. We calculated the SOC magnitude between the driving singlet excited state and nearby triplet electronic states with the SA3-CASSCF(8/8)/aug-cc-pVDZ method along our LIICs (see SM) and observed that the SOC magnitude remains smaller than 20~cm$^{-1}$ in the vicinity of the S$_2$ minimum. Such a value of SOC would require the photoexcited molecule to exhibit an extended lifetime in the S$_2$ electronic state, i.e., picoseconds,\cite{favero2013dynamics} to undergo an efficient deactivation via ISC. By analogy with the photodynamics of cyclobutanone excited in the S$_1$ electronic state, a cyclobutanone with a long-lived S$_2$ state can be prepared by photoexcitation below the dissociation barrier. In the S$_1$ state, excitation above 315~nm (red edge of the first band in the photoabsorption cross-section) creates a photoexcited molecule with internal energy below the dissociation barrier, opening the ISC channel for deactivation. We note that the 200~nm laser pump that will be used to photoexcite cyclobutanone will also target the red edge of the photoabsorption cross-section (see Fig.~\ref{fig:spec}a). Therefore, a similar scenario could be plausible for the S$_2$ ring-opening barrier, as the S$_2$ barrier height is 0.145 eV (both with and without zero-point energy, ZPE) and comparable to that of the S$_1$ state, 0.176 eV without ZPE and 0.146 with ZPE. Thus, an ISC channel might be operational on a long timescale during the planned MeV-UED experiment, hampering the IC process. However, earlier experimental studies\cite{Kuhlman2012a} at 200~nm (and our simulations presented below) reveal an ultrafast internal conversion of cyclobutanone from the S$_2$ electronic state and cast doubt on the ISC hypothesis.
We finally note that for the ring-opened S$_1$ structure, the SOC magnitude between the S$_1$ state and neighboring triplet states rises up to 40~cm$^{-1}$ yet we will show below that the lifetime of this species in S$_1$ is again too short for ISC to occur. 

\subsubsection{Steady-state UED signals of cyclobutanone}

In the following, we offer a first prediction of a quantity that can be retrieved from the MeV-UED experiment, the pair-distribution function (PDF), on the support of the two reaction pathways we discussed earlier. 

\begin{figure}[ht]
    \centering
    \includegraphics[width=0.7\textwidth]{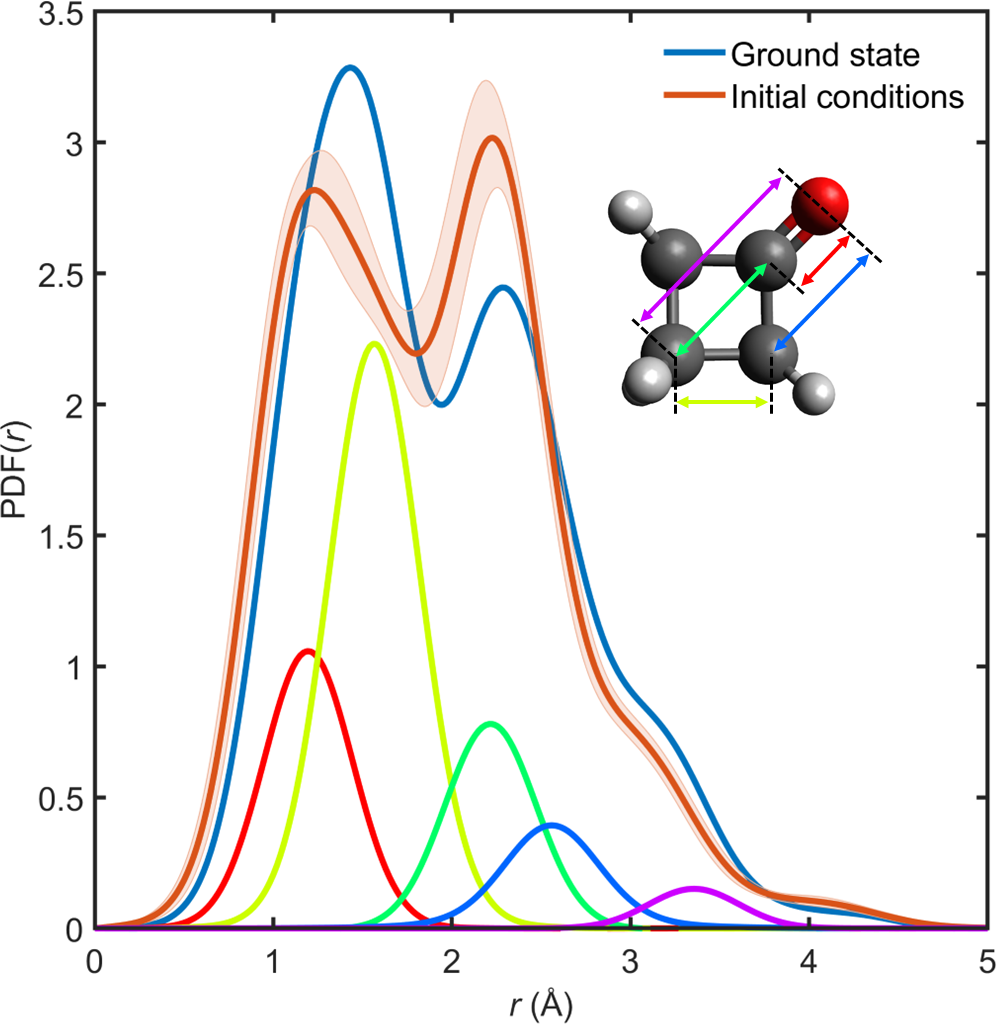}
    \caption{Steady-state PDF of cyclobutanone. The PDF was calculated for the ground-state minimum (XMS-CASPT2(8/8)/aug-cc-pVDZ) and as an average of the PDF calculated for each initial condition (119) used for the nonadiabatic dynamics (sampled from the QT-BOMD trajectory, MP2/cc-pVDZ). The PDF signal obtained from the initial conditions is decomposed into different atom-pair distances relevant to cyclobutanone, highlighted in the molecular structure given as inset. Note that the PDF signal depicted by the lime curve corresponds to all the \ce{C-C} bonds of cyclobutanone.}
    \label{fig:pdf_static}
\end{figure}

Let us begin by inspecting the PDF for the ground-state cyclobutanone (Fig.~\ref{fig:pdf_static}) resulting from the ground-state optimized geometry and that obtained as an average over the PDFs for each initial condition (used to initiate our nonadiabatic molecular dynamics). We note that the PDF acquired for the 119 initial conditions is in close agreement with the PDF obtained from the 5000 geometries sampled from the QT-BOMD trajectory and, therefore, is believed to offer a reliable depiction of a PDF for the S$_0$ nuclear probability density of cyclobutanone, see SM. First, we decompose the average PDF obtained for the initial conditions into contributions of different atom pairs observed in the molecule (see Fig.~\ref{fig:pdf_static} and the molecular structure for the color code). The first peak of the overall PDF for the initial conditions (centered around 1.5~{\AA}) corresponds to bonded pairs of atoms -- the shortest distance comes from the carbonyl bond (red curve in Fig.~\ref{fig:pdf_static}), while the main contribution (lime curve) originates from all the \ce{C-C} bonds of the molecular framework. The second peak of the main PDF for the initial conditions, centered at around 2.5~{\AA}, is a combination of non-bonded atom pairs, mainly composed of distances between the carbon or oxygen of the carbonyl with the three opposite carbons in the cyclobutanone ring. Interestingly, while the second peak of the PDF for the ground-state optimized geometry exhibits a lower intensity than the first peak, this order is reversed in the PDF obtained from the whole set of initial conditions. As stressed before, the ground-state minimum is not fully representative of the ground-state nuclear probability density, and the non-bonded distances appear to play a more significant role when the nuclear density is considered. Thus, care needs to be taken when using optimized geometries to analyze an experimental UED signal, and adequate nuclear probability density should be considered instead (reinforcing the message of Ref.~\citenum{nunes2023monitoring}). 

\begin{figure}[ht]
    \centering
    \includegraphics[width=1.0\textwidth]{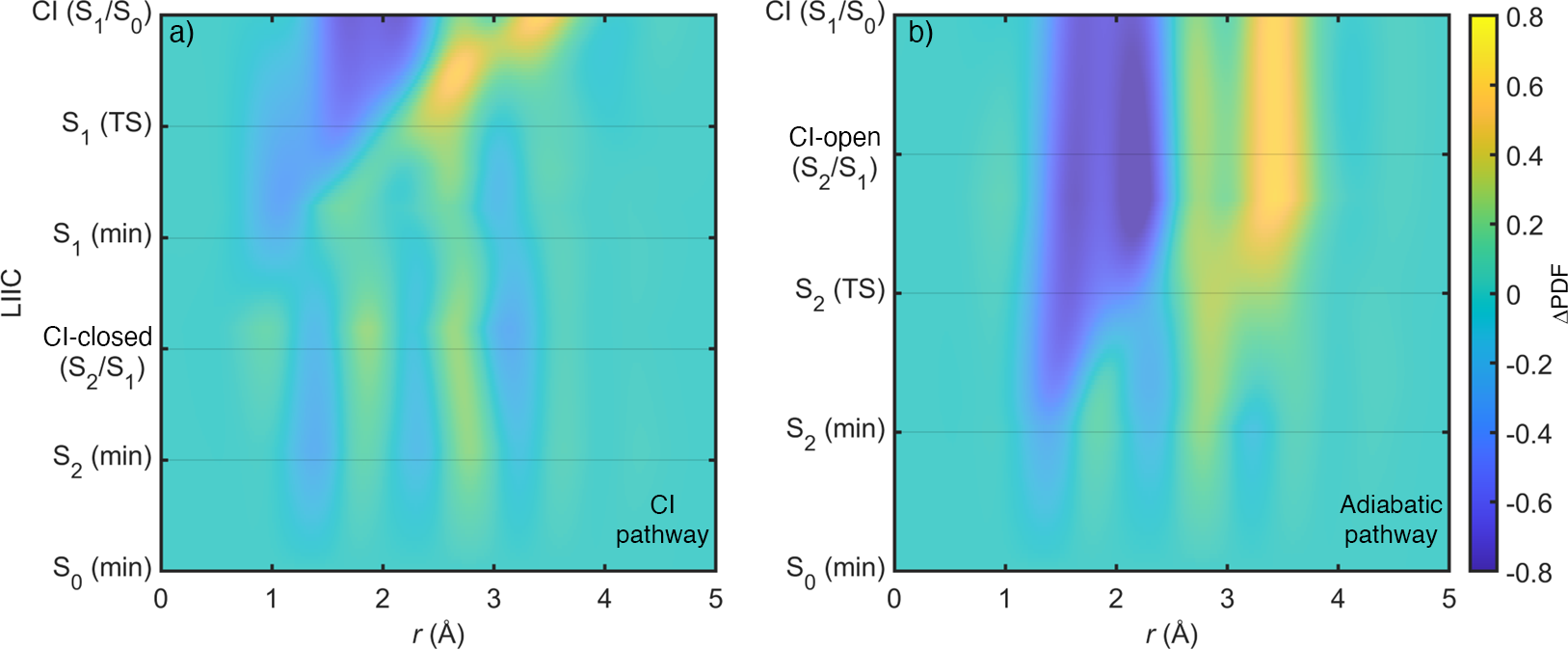}
    \caption{Simulated $\Delta$PDF signal (PDF for a given geometry along the LIIC minus the PDF of the reference S$_0$ minimum geometry) along the two deactivation pathways -- a) the CI pathway and b) the adiabatic pathway -- of cyclobutanone excited at 200 nm. The geometries used to predict $\Delta$PDF signals are the same as those used for the electronic energies presented in Fig.~\ref{fig:liic}, with the only exception of the transition-state geometries (TS) along the CI pathway that was added for a more detailed analysis of the ring-opening process.}
    \label{fig:pdf_liic}
\end{figure}

Let us now move to a more conceptual analysis of our data by predicting the PDF signal along the two LIIC pathways discussed in Fig.~\ref{fig:liic}. Needless to stress that such an analysis needs to be taken with a grain of salt based on the previous paragraph as it solely relies on critical geometries and interpolated geometries between them -- most likely not representative of a time-resolved signal that we will discuss later based on nonadiabatic molecular dynamics simulations. The difference PDFs ($\Delta$PDFs, obtained as the PDF for a given geometry on the LIIC minus the PDF of the reference S$_0$ minimum geometry) for both LIICs introduced earlier are presented in Fig.~\ref{fig:pdf_liic}. The initial step of both pathways is the relaxation of cyclobutanone from its S$_0$ minimum geometry to the S$_2$ minimum geometry, which manifests itself as a fringe of positive and negative $\Delta$PDF signals caused by the elongation of all distances between the carbonyl group and the other carbon atoms due to planarization. This elongation leads to negative $\Delta$PDF signals for each of the original S$_0$ minimum atom-pair distances (shown in Fig.~\ref{fig:pdf_static}) and positive $\Delta$PDF signals for the now increased atom-pair distances, hence the appearance of fringe on the overall $\Delta$PDF plot for this part of both pathways.

We shall now focus on the CI pathway. The fringe pattern persists during the passage through CI-closed(S$_2$/S$_1$) and until cyclobutanone reaches the S$_1$ minimum. The only difference in this part of the path appears around 1~{\AA}. This change is caused by the \ce{C=O} bond that first shortens from 1.17~{\AA} to 1.12~{\AA} when going from the S$_2$ minimum to the CI-closed(S$_2$/S$_1$) -- explaining the positive $\Delta$PDF signal observed near 1~{\AA} around CI-closed(S$_2$/S$_1$). The \ce{C=O} bond then stretches to 1.30~{\AA} when reaching the S$_1$ minimum, consistent with the $n\pi^*$ character of this state and leading to a depletion of the $\Delta$PDF signal at the original S$_0$ \ce{C=O} bond distance and the appearance of a negative $\Delta$PDF feature at around 1.20~{\AA}.
The following ring-opening process over the transition state leads to a substantial depletion of the $\Delta$PDF signal around 2~{\AA} and a connected increase around 3~{\AA} when the molecule reaches the CI(S$_1$/S$_0$), clearly attesting from the loss of a bonded interaction and shorter non-bonded atom-pair distances, and the appearance of further (non-bonded) distances from the ring-opened structure. Focusing on the adiabatic pathway, we observe a very similar pattern to that of the CI pathway when the cyclobutanone ring opens, simply arriving earlier in the deactivation pathway as the S$_2$ TS already results in a ring opening. The overall $\Delta$PDF pattern barely changes between CI-open(S$_2$/S$_1$) and CI(S$_1$/S$_0$) as both geometries are open and rather similar geometrically. We note that the evolution of different bond distances depicted in the inset of Fig.~\ref{fig:pdf_static} is presented in the SM for more insight on the $\Delta$PDFs along the LIICs.

The differences between the pathways do not appear to be very pronounced, and distinguishing between them in the experiment might be difficult. The evolution of the photoexcited cyclobutanone on S$_2$ is likely to be visible, and an interesting question will be the time the molecule spends near the S$_2$ minimum before finding its way to the CI-closed(S$_2$/S$_1$) or the S$_2$ TS. The vibronic structure observed in the photoabsorption cross-section (Fig.~\ref{fig:spec}) and the broadening of its peaks seem to indicate that cyclobutanone should at least remain in this region of configuration space for a few hundred femtoseconds, which is corroborated by earlier time-resolved photoelectron spectroscopy and time-resolved mass spectroscopy experiments.\cite{Kuhlman2012a} While the ring opening of cyclobutanone will be clearly visible when it happens, both pathways -- ring opening in S$_1$ or S$_2$ -- lead to very similar features in the $\Delta$PDF. Probably the most significant difference between the pathways lies in the variation of the $\Delta$PDF signal for molecular structures near the CI-closed(S$_2$/S$_1$) and S$_1$ minimum. While the CI-closed(S$_2$/S$_1$) will only act as a gateway  (the nuclear wavepacket will only visit this part of configuration space for a short amount of time), the S$_1$ minimum could be sufficiently long-lived to be observed. The S$_1$ minimum is characterized by an extended \ce{C=O} bond and a shorter distance between the carbonyl group and the opposite carbon. As a result, the fringe is shifted to shorter distances compared to the S$_2$ minimum (see Fig.~\ref{fig:pdf_liic}a).

\subsection{Time-resolved photochemistry of cyclobutanone}
\label{timeresolved}

\subsubsection{Analysis of the nonadiabatic molecular dynamics simulations}

\begin{figure}[ht]
    \centering
    \includegraphics[width=\textwidth]{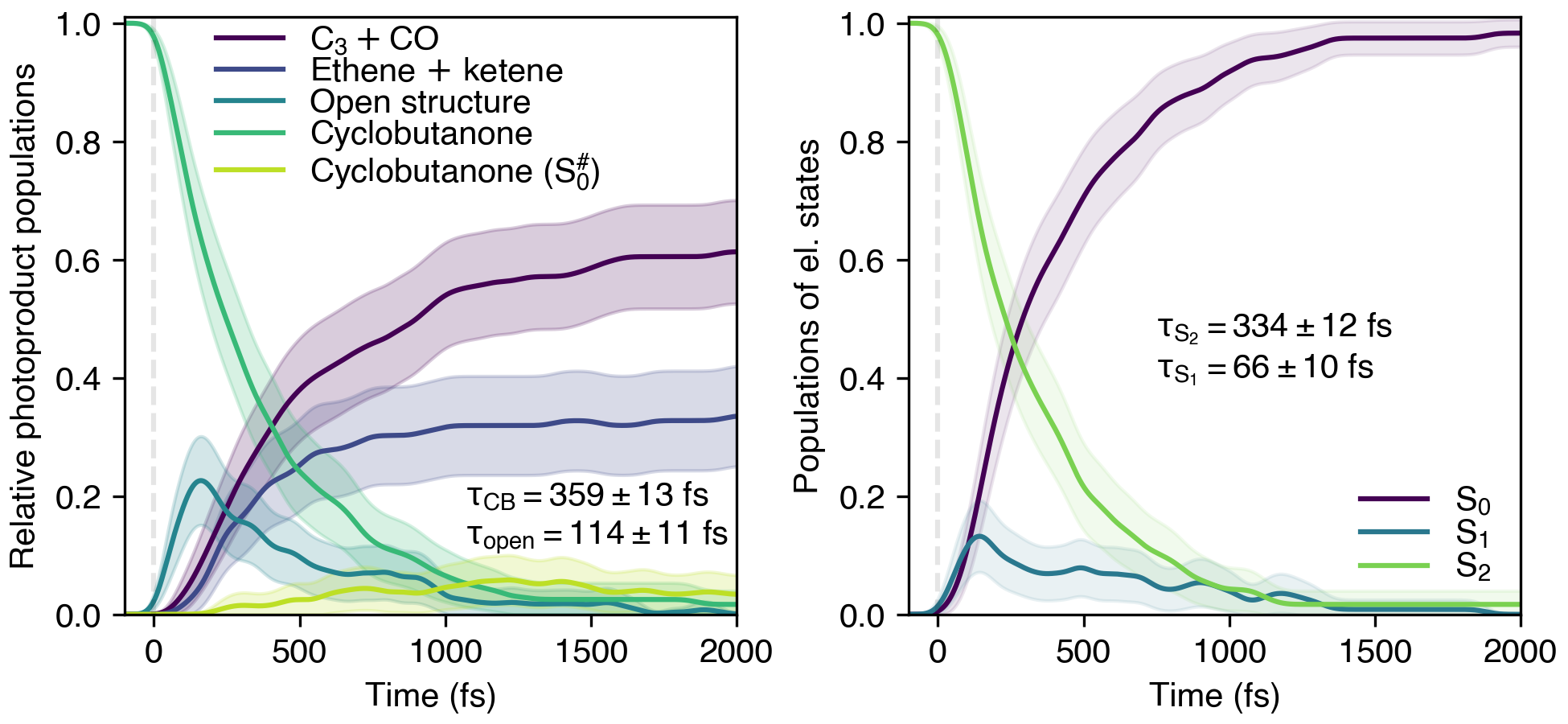}
    \caption{Relative populations of the photoproducts (left) and the electronic states (right) obtained from the (NA+BO)MD for a swarm of 119 trajectories. The population traces were convoluted with the envelope of the laser pulse. The values report the fitted lifetimes for the population of cyclobutanone in an excited electronic state ($\tau_{\mathrm{CB}}$) and the open structure ($\tau_{\mathrm{open}}$), as well as the electronic population of the S$_2$ state ($\tau_{\mathrm{S_2}}$) and the S$_1$ state ($\tau_{\mathrm{S_1}}$). The details on the fitting procedure and resulting curves are described in the SM.}
    \label{fig:analysis}
\end{figure}

We begin this discussion of our (NA+BO)MD simulations of cyclobutanone, photoexcited in its S$_2$ electronic state, by inspecting the resulting populations predicted by the swarm of trajectories. Let us first focus on the time-dependent relative populations of the cyclobutanone and photoproducts (left panel of Fig.~\ref{fig:analysis}). The population of the parent cyclobutanone molecule rapidly decays (lifetime of $\sim$359~fs) following photoexcitation due to a ring-opening induced by the cleavage of one of the \ce{C-C} bonds adjacent to the carbonyl group. This open structure acts as an intermediate to the formation of either CO + C$_3$ (cyclopropane or propene) or ethene + ketene (C$_2$), or the reformation of a hot ground-state cyclobutanone with a ratio of 71:42:4. The C$_3$ products observed are dominated by cyclopropane (83\%), followed by propene (17\%). The formation of cyclopropane or propene is mediated by the appearance of an intermediate \ce{^{.}CH2-CH2-CH2^{.}} biradical structure.  We did not observe the conversion of cyclopropane to propene (or \textit{vice versa}) in our simulation, i.e., the propene was always formed directly from the photochemical reaction and not later in the ground state via the isomerization from cyclopropane, as suggested in the literature. For the C$_2$ channel, we observed the decomposition of ketene into \ce{CH2} and CO in the ground electronic state as a minor channel (only 2 trajectories). A small population of hot ground-state cyclobutanone (S$_0^\#$) is reformed during the photochemical process, yet it was observed to be unstable and decomposed into one of the other photoproducts after several hundreds of femtoseconds. The relative populations of the photoproducts after 2~ps are summarized in Table~\ref{tab:qy}. 

\begin{table}[ht!]
    \centering
    \begin{tabular}{l r}
        \hline
        photoproducts & relative population in \% \\
        \hline
         CO + cyclopropane & $ 49.6 \pm 9.0$ \\
         CO + propene & $ 10.1 \pm 5.4$ \\
         CO + biradical & $ 0.0 \pm 0.0 $ \\
         ethene + ketene & $ 33.6 \pm 8.5 $ \\
         ethene + \ce{CO} + \ce{CH2} & $ 1.7 \pm 2.3 $ \\
         cyclobutanone & $ 1.7 \pm 2.3 $ \\
         cyclobutanone(S$_0^\#$) & $ 3.4 \pm 3.2 $ \\
         \hline
    \end{tabular}
    \caption{Relative populations of the photoproducts of cyclobutanone following irradiation at 200~nm as predicted after 2~ps of (NA+BO)MD. We note that these populations vary over time and have to be considered as a snapshot of the photoproduct populations in the earlier times following the photoexcitation. }
    \label{tab:qy}
\end{table}

The decay of the electronic-state populations exhibits similar time scales, with the lifetime of the S$_2$ population predicted to be $\sim$334~fs and that of the S$_1$ state $\sim$66~fs. While the majority of the trajectories proceeded via the expected cascade $\mathrm{S}_2\rightarrow\mathrm{S}_1\rightarrow\mathrm{S}_0$, 12 out of the 119 trajectories hopped directly from S$_2$. to S$_0$ These hops always happened during the adiabatic pathway close to the CI-open geometry (see Fig.~\ref{fig:liic}), where all three electronic states are coupled, and their respective FSSH electronic coefficients are non-zero.

From a mechanistic perspective, our (NA+BO)MD trajectories indicate that both the adiabatic and CI pathways are operative, with the adiabatic pathway being the dominant channel. A detailed analysis of the hopping geometries between the S$_2$ and S$_1$ states using a multidimensional scaling technique indeed reveals that the hops through the CI-open (adiabatic pathway) versus those through the CI-closed (CI pathway) follow a 3:1 ratio (see SM for additional details). Apart from these two pathways, we observed a few additional minor channels: the molecule can dissociate into ethene and ketene in the S$_1$ electronic state and then transfer to the ground state, or the molecule can remain open in the ground state for a few hundreds of femtoseconds before dissociating and forming one of the products.

Finally, we note that we also ran 45 trajectories using an LZSH procedure (using the same level of electronic-structure theory) to probe the sensitivity of the dynamics to the nonadiabatic dynamics algorithm. We did not observe any qualitative discrepancy for the 45 trajectories within 700~fs of dynamics, reinforcing our confidence in the FSSH trajectories. We also ran a set of 10 LZSH/SA(3S,3T)-CASSCF(8/8) trajectories, including the triplet states, to confirm that the ISC pathway should be minor for photoexcitation of cyclobutanone to its S$_2$ electronic state, see SM for more details.

\subsubsection{Time-resolved MeV-UED signal}

\begin{figure}[ht]
    \centering
    \includegraphics[width=\textwidth]{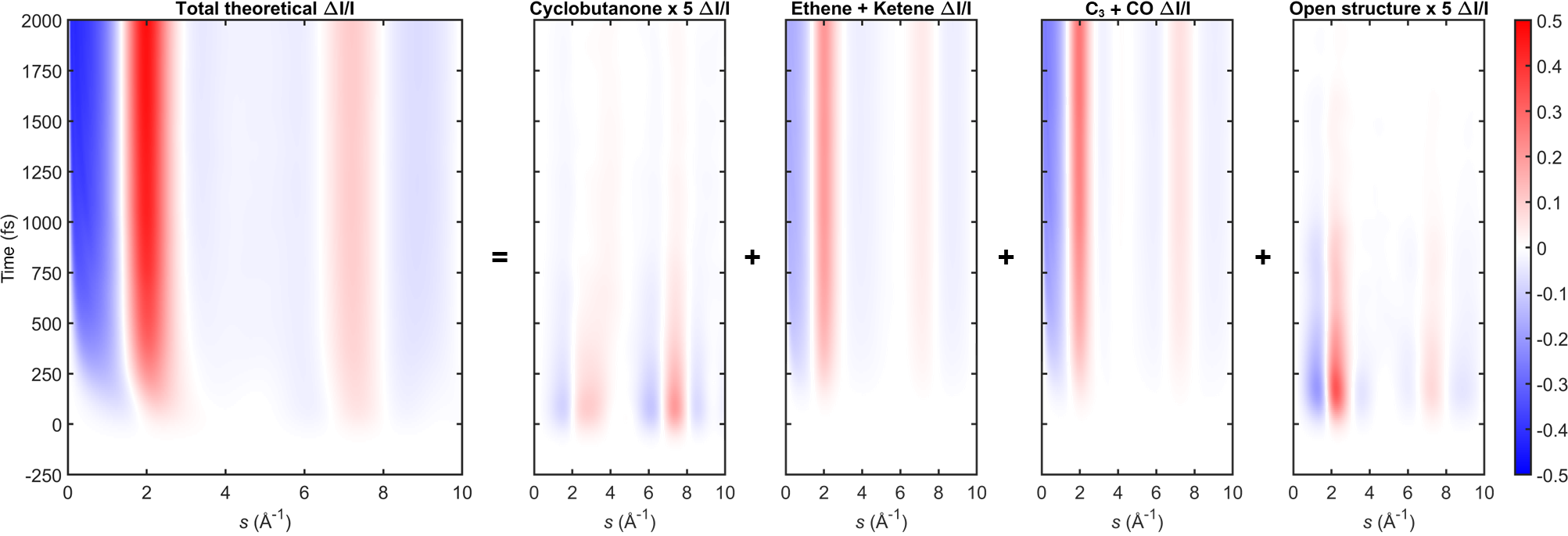}
    \caption{Time-resolved $\Delta I /I(s,t)$ MeV-UED signal obtained from the swarm of (NA+BO)MD trajectories and convoluted with the instrument response function. The total $\Delta I /I(s,t)$ signal is presented on the left panel and then decomposed into the contribution of cyclobutanone and each category of photoproducts. }
    \label{fig:dioveri_decomp}
\end{figure}

We present here the prediction of time-resolved and time-averaged MeV-UED signals based on the swarm of (NA+BO)MD trajectories. The $\Delta I/I(s,t)$ signal (Fig.~\ref{fig:dioveri_decomp}) was evaluated from the molecular structures along the (NA+BO)MD trajectories using the IAM and convoluted with the instrument response function (150~fs FWHM\cite{nunes_mev_ued}). We note that this convolution smears out some fine details of the signal, like the early-time oscillations produced by the photoexcited cyclobutanone in the S$_2$ electronic state. The total $\Delta I/I(s,t)$ signal (left panel of Fig.~\ref{fig:dioveri_decomp}) only exhibits weak variations over the course of time, in particular after 500 fs, and displays two strong positive  ($\sim$ 2.5 and 7.5~$\mathrm{\AA}^{-1}$) and negative (between 0 and 1.5~$\mathrm{\AA}^{-1}$) features. The rest of the $s$ range considered is mostly composed of weak negative contributions. As introduced in Ref.~\citenum{nunes2023monitoring}, we exploited one advantage of theory over experiment: the total $\Delta I/I(s,t)$  can be decomposed into the contributions of each different species (right panels of Fig.~\ref{fig:dioveri_decomp}), providing an alternative interpretation tool for the features observed in the total signal. From this decomposition, we can directly observe that the contribution of the photoproducts (ethene + ketene and C$_3$ + CO) to the $\Delta I/I(s,t)$ signal is about 5 times stronger than that of cyclobutanone or its open structure and explain the strong features highlighted before. 

\begin{figure}[ht]
    \centering
    \includegraphics[width=\textwidth]{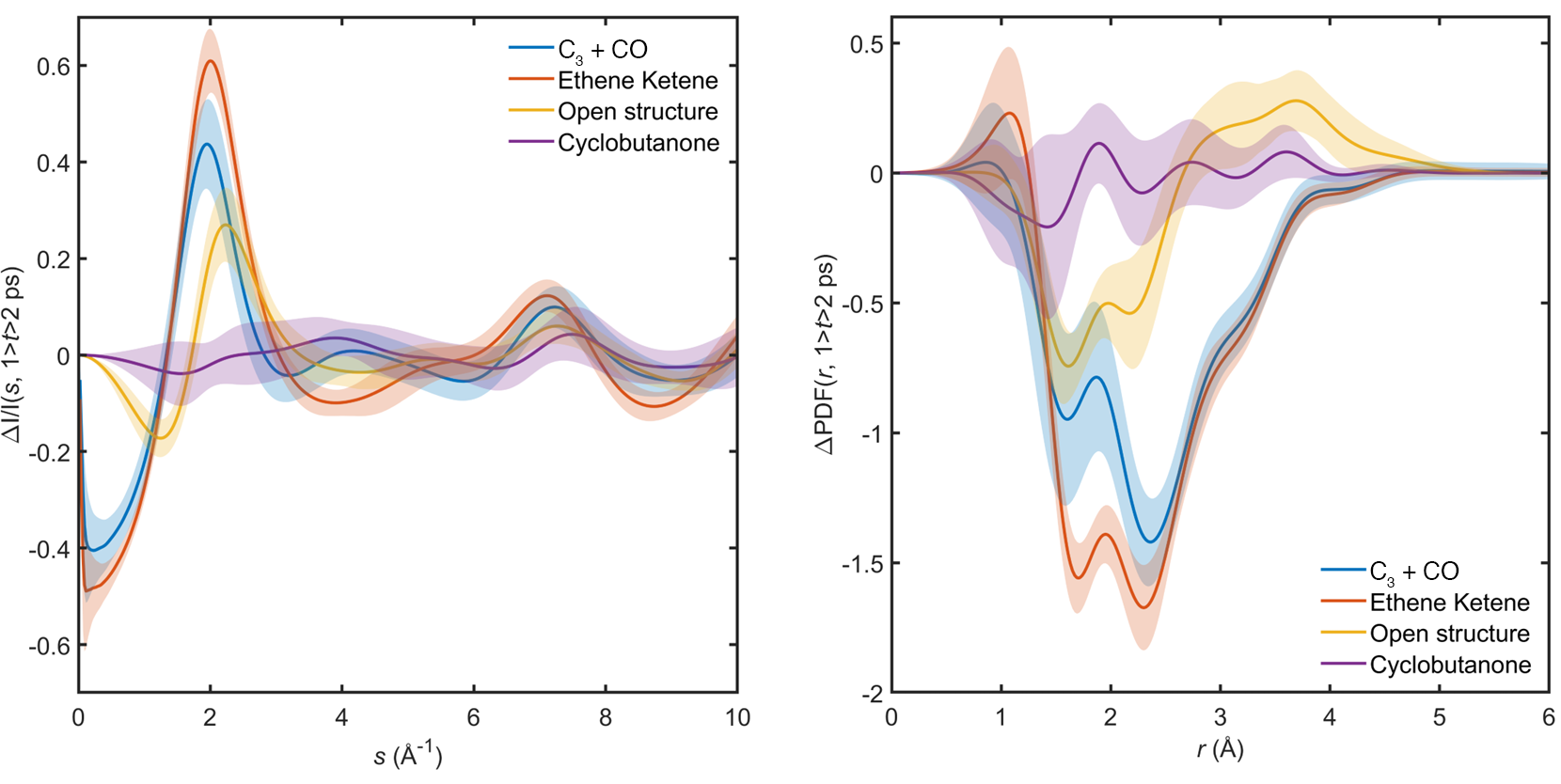}
    \caption{Representative $\Delta I /I(s)$ (left) and $\Delta$PDF$(r)$ (right) signals for cyclobutanone and the different families of photoproducts, obtained from the averaging of the theoretical $\Delta I /I(s,t)$ and $\Delta$PDF$(r,t)$ signals for each category of molecules in the time interval 1 - 2 ps. }
    \label{fig:dpdf_decomp}
\end{figure}

Given the rather time-independent nature of the decomposed $\Delta I /I(s,t)$ UED signals in the region between 1 and 2~ps, we average them to produce representative $\Delta I /I(s)$ signals for each category of photoproducts and cyclobutanone (left panel of Fig.~\ref{fig:dpdf_decomp}). These signals can be used as time-independent basis functions to fit the corresponding experimental signal and decompose it into the contributions of the main different molecular structures (as performed in Ref.~\citenum{nunes2023monitoring}). The features (amplitudes and locations of the peaks) of each time-independent $\Delta I /I(s)$ basis function produced are different enough to be used in a fitting procedure. We note that distinguishing between the different C$_3$ products (cyclopropane and propene) would most likely not be possible experimentally due to experimental resolution -- this observation motivated the categories proposed for our general analysis of the (NA+BO)MD results. We also deployed the same procedure to produce averaged $\Delta$PDF$(r)$ for the photoproducts and cyclobutanone -- these signals being easier to interpret in terms of molecular parameters as discussed later.

\begin{figure}[ht]
    \centering
    \includegraphics[width=0.6\textwidth]{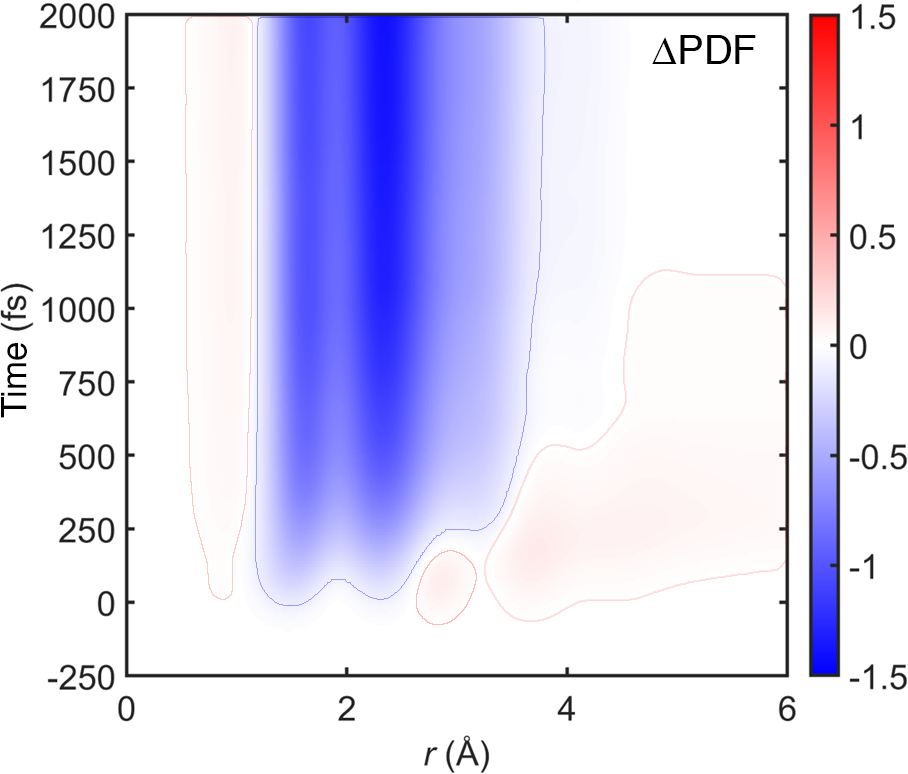}
    \caption{Time-resolved $\Delta$PDF$(r,t)$ signal obtained from the swarm of (NA+BO)MD trajectories and convoluted with the instrument response function. Contour lines around the signal map were added to improve the visibility of weak features.}
    \label{fig:dpdf}
\end{figure}

Let us finally move to the interpretation of the predicted time-resolved $\Delta$PDF$(r,t)$ signal calculated from our (NA+BO)MD trajectories (Fig.~\ref{fig:dpdf}) by using the basis functions introduced above (right panel of Fig.~\ref{fig:dpdf_decomp}). Focusing on the short distances, we observe a constant weak positive signal at around 1~$\mathrm{\AA}$ from the early times of the dynamics, corresponding to a loss of single carbon bonds in cyclobutanone and formation of double bonds in ethene and ketene (see the basis function for ethene + ketene in Fig.~\ref{fig:dpdf_decomp}). Centered at around 2~$\mathrm{\AA}$, we observe two strong negative features caused by a loss of chemical bonds during dissociation. This loss can be equally attributed to the dissociation into C$_2$ and C$_3$ photoproducts or the open structure, see Fig.~\ref{fig:dpdf_decomp}. While the described signal below 2.5~$\mathrm{\AA}$ is constant throughout the entire dynamics, the region above 2.5~$\mathrm{\AA}$ shows time-dependent features. Within the first 250~fs of dynamics, two positive signals between 3 and 4~$\mathrm{\AA}$ appear due to the formation of the open structure, which is consistent with its relative population calculated in Fig.~\ref{fig:analysis}. This positive signal appears then to spread to larger distances $r$ as the fragments dissociate, and the signal in the region between 3 and 4~$\mathrm{\AA}$ becomes negative due to the formation of the photoproducts. Since the positive signal characteristic of the dissociation is low, we expect that a direct observation of the dissociation features on a long timescale might be challenging. With that said, Figure~\ref{fig:dpdf_decomp} shows that dissociation produces a very strong feature at low $s$, which can be clearly identified in the UED signal, even at later times. To distinguish between dissociation and ring-opening, it would be important that the low $s$ signal ($0.5>s>1$) is recorded. 

We note that the fringe pattern representative of the S$_2$ minimum predicted by our static calculations along the LIICs can be visible only in the unconvoluted signal during the first 40~fs. Nevertheless, it's quickly overlayed by the ring opening and the convolution with the instrument response function smears it out. Still, the fringe pattern might be visible in the experiment considering the potential limitations of our simulations (see Sec.~\ref{sec:limitations} below), as the ring-opening of cyclobutanone in the excited state may be too rapid in our simulations (due to the accuracy of the electronic-structure method and the internal energy of the photoexcited cyclobutanone). 

\section{Limitations of the current model and their impact on the predicted experimental signals}
\label{sec:limitations}

In this Section, we summarize the limitations of the model used for our predictions, arising either from the approximations underlying our methodology or the time constraint imposed by this prediction challenge. We grouped the limitations into four categories corresponding to the various components of our modeling: level of electronic-structure theory, the influence of initial conditions, approximations of the nonadiabatic molecular dynamics, and generation of the UED signal. We also discuss the implications of these limitations on the predictions of the experimental signals and the corresponding \textquote{confidence intervals}.

\subsection{Electronic-structure theory}

The electronic structure has been recognized as one of the most quality-determining components of nonadiabatic molecular dynamics simulations. We selected the XMS-CASPT2 method for our simulations due to \textit{i}) the homolytic bond breaking that requires a multiconfigurational method and \textit{ii}) the sensitivity of the activation barriers for dissociation to dynamic correlation observed in our LIIC. All the other methods at hand (LR-TDDFT, ADC(2), SI-SA-REKS, SA-CASSCF, or FOMO-CASCI) were either not compatible with a homolytic dissociation or would not provide an accurate depiction of the energy barriers along the dissociation profile. Nevertheless, the performance of XMS-CASPT2 strongly depends on the choice of its parameters, such as the active space. We offered a detailed benchmark to justify our methodology, but with more time at hand, we would have investigated further the following potential improvements of our XMS-CASPT2 calculations. 

(a) Inclusion of the two $\sigma_\mathrm{CC}$ and corresponding two $\sigma^*_\mathrm{CC}$ orbitals opposite to the carbonyl group, which facilitate the formation of the ketene and ethene photoproducts. Our active space does not include these orbitals from the beginning, and our simulations rely on (slowly) rotating these orbitals in the active space once the adjacent \ce{C-C} bond is open and its corresponding $\sigma$ orbitals become irrelevant. Therefore, the formation of ketene and ethene depends on uncontrolled orbital rotations and can be hampered compared to the release of CO, for which all the relevant orbitals are present in the active space from the beginning of the dynamics. This extended (12/12) active space would require more computational time than this challenge allowed and was, as such, deemed out of reach. 

(b) Inclusion of higher excited electronic states in the state averaging to stabilize the active space and prevent jumps in total energy. The issues with energy discontinuities in our simulations partially stemmed from the S$_2$ electronic state increasing in energy upon dissociation (due to the character of this electronic state), leading eventually to a root flipping. However, as for point (a) above, this increased state-averaging protocol would require the inclusion of $Ry(3p)$ orbitals in the active space, meaning an even larger active space and prohibitive computational costs for this challenge. 

(c) Inclusion of the triplet electronic states in the state averaging. This inclusion would allow us to monitor the influence of SOC along the trajectories (more rigorously than our tests with LZSH). However, the (8/8) active space used in this work is unstable with respect to the T$_2$ ($\pi\pi^*$) electronic state, which quickly rises in energy and swaps with other triplet electronic states, causing a jump in total energy.

\subsection{Influence of initial conditions}

The selection of initial conditions in photodynamics simulations is a topic of intense research in the community and the modeled experiment represents a challenging (and exciting) case in this context. 
Our modeling relies on a sudden vertical excitation of the molecule, projecting the ground state on the S$_2$ excited electronic state, combined with an energy-windowing scheme, that accounts for the energy spread of the laser pulse. 
The main limitations of this approach are:

(a) The vertical excitation approximation is unable to correctly describe excitation to well-defined vibrational states, which is expected with the experimental pump laser pulse.
Although we used the windowing scheme, our simulations still remain within the frame of a sudden excitation of the molecule, which is justified only for a $\delta$ laser pulse. As a result, the spread of total energy for our simulated nuclear wave packet at the beginning of the nonadiabatic dynamics is about 0.84 eV, while it should be around 0.05 eV for the experimental pulse. 
This observation is crucial if one considers that the photoexcited cyclobutanone has to cross a barrier following photoexcitation to the S$_2$ electronic state. The excess of internal energy in our initial nuclear wave packet may result in a shorter lifetime of the excited molecule in S$_2$ and hamper the ISC channel.

(b) The sudden approximation used in this work leads to the formation of a perfect nuclear wavepacket in S$_2$, meaning that the experimental signal simulated is most likely too coherent. A more suitable generation of initial conditions can be achieved by a technique developed by Saalfrank et al,\cite{Martinez-Mesa2015} which includes the laser pulse explicitly to provide time-dependent initial conditions. However, an implementation of this strategy for molecular systems is currently not available.

(c) The ground-state nuclear density was sampled at 298 K, as the true temperature after a nozzle expansion was not clearly defined.

\subsection{Approximations of the nonadiabatic molecular dynamics strategy}

Our results are based on FSSH trajectories that were benchmarked against the simpler LZSH technique. Although both methods yielded similar population traces, several limiting factors of this trajectory-based approach need to be highlighted.

(a) FSSH may suffer from its independent trajectory approximation. We observed for some of our trajectories that the FSSH electronic coefficients were non-zero for tens of femtoseconds following nonadiabatic transitions. The decoherence is accounted for via the \textit{ad hoc} energy-based decoherence correction in the present study, yet this correction appears to compete with the transfer of electronic amplitude during the dynamics through the nonadiabatic regions. This observation calls for additional tests on the reliability of decoherence-corrected FSSH, for example by comparing its result with that obtained with a multiple-spawning based approach like AIMSWISS.

(b) The three-state hops observed in our dynamics (12 out of 119 trajectories) might also pose a challenge for the FSSH algorithm, and further benchmarking with a multiple-spawning-based approach like AIMSWISS would be desirable.

(c) The simulations do not incorporate ISC processes, rationalized by our static calculations and tests using LZSH/CASSCF. It would be safer to monitor the magnitude of SOC along the trajectories and preserve the possibility for ISC channels. As stated above, this approximation is mostly motivated by the availability and cost of the electronic structure and the potential need for a large number of trajectories to describe ISC adequately.\cite{persico2014overview} 

(d) Zero-point energy leakage can affect our semiclassical dynamics on the 2~ps timescale.\cite{doi:10.1021/acs.jctc.3c00024} As a result, our lifetimes are likely to be shorter than predicted by quantum dynamics.

\subsection{Generation of the UED signal}

Our prediction can also be affected by the IAM applied to approximate the UED signal from our FSSH trajectories. Although the effect of the electronic charge density is often negligible, it was shown that it could play a significant role in the evaluation of signal.\cite{Champenois2023} 

\subsection{Direct implications of these limitations on the predicted MeV-UED signals}

Based on the limitations highlighted below, we would expect to see a longer lifetime of cyclobutanone in its photoexcited S$_2$ electronic state. The time-dependent relative populations of the photoproducts obtained should be considered with larger error bars than the statistical ones provided. We expect that the active space used favours the CO release over the formation of ethene and ketene, although other effects also might play a role. The ratio of photoproducts and lifetimes predicted might also differ because the pump laser pulse will most likely form the excited-state wave packet with lower internal energy than our sudden excitation estimate, which may appear at the edge of the dissociation barrier and CI-closed. If this happens, we would then expect the MeV-UED signal representative of the S$_2$(min) to last for longer and, possibly, to observe an alternative ISC channel on a scale of a few to tens of picoseconds. The ISC channel would dominantly form CO and cyclopropanone/propene based on the experiments with 340~nm pump pulse\cite{Tang1976}. We note, however, that a previous experiment using the same excitation wavelength observed subpicosecond deactivation processes.\cite{Kuhlman2012a} 

\section{Conclusion}

In this work, we investigated the first two picoseconds of the cyclobutanone photodynamics following an excitation with a laser pulse centered at 200~nm and predicted the expected time-resolved MeV-UED signal for the processes observed. Our (NA+BO)MD trajectories show an almost complete return of the nuclear wavepacket to the ground state within the timescale of our simulations. The main photoproduct observed from this photochemical reaction after 2~ps is CO (60\%) accompanied by either cyclopropane (50\%) or propene (10\%), while the second photoproduct was ethene and ketene (34\%). An additional minor product was detected: ethene with CO and \ce{CH2}, which is formed by a dissociation of the ground-state ketene and a hot ground-state cyclobutanone. The formation of photoproducts follows two main pathways. In the adiabatic pathway, the photoexcited cyclobutanone ring opens next to the carbonyl group in the S$_2$ electronic state to form an open biradical structure, which then decays rapidly via internal conversion to the ground electronic state S$_0$, where the photoproducts are formed. In the CI pathway, the photoexcited cyclobutanone undergoes an internal conversion to S$_1$ through a conical intersection characterized by a closed cyclobutanone structure. The \ce{C-C} bond next to the carbonyl group is then cleaved and photoproducts are formed upon internal conversion to the S$_0$ state. Our (NA+BO)MD trajectories predict that the ratio of the adiabatic to CI pathways is approximately 3:1. The decomposition of cyclobutanone is characterized by a lifetime estimated to be $359\pm13$~fs, which correlates with the S$_2$-state population lifetime of $334\pm12$~fs. Similarly, the lifetime of the ring-open structure ($114\pm11$~fs) correlates with the lifetime of the population in the S$_1$ state ($66\pm10$~fs).

The main outcome of this work is the prediction of a series of MeV-UED signals for the photochemistry of cyclobutanone excited at 200~nm. A stationary MeV-UED signal ($\Delta$PDF) was predicted along interpolated paths in internal coordinates, characterizing the critical structures relevant for the adiabatic and CI pathways. Our swarm of (NA+BO)MD trajectories was also used to predict time-resolved MEV-UED signals ($\Delta$PDF and $\Delta I/I$) that would be observed for the 2~ps nuclear wavepacket dynamics simulated in our work. We decomposed the $\Delta I/I$ signal into its molecular components (cyclobutanone and the different photoproducts observed) and suggested time-independent basis functions -- based on the average, athermal structures observed in the dynamics -- that could be used to distinguish between the following molecules when they are evolving in the hot ground state: hot cyclobutanone, C$_3$ products + CO, ethene + ketene, and the ring-open structure.

We further stress the challenges one faces when predicting such experiments and the expected limitations of our approaches, which are difficult to overcome either due the current limitations of the methodologies or the time constraints imposed by this prediction challenge. As emphasized in recent work,\cite{Janos2023} the level of electronic-structure theory appears to play a decisive role in the quality of the predictions, both stationary and time-resolved ones, given in particular the sensitivity of the excited-state dissociation barrier in the S$_2$ electronic state to the quantum-chemical method used and its importance in controlling the efficiency of the ring-opening process. This dissociation process appears to require both a multiconfigurational treatment and a proper inclusion of dynamic correlation, leading us to employ the XMS-CASPT2 method, which, however, is highly sensitive to the precise selection of its underlying active space and the number of electronic states considered, leading to very computationally-demanding simulations. The price to be paid for using XMS-CASPT2 forced us to compromise and combine it with the mixed quantum/classical method FSSH, which may not offer the most accurate depiction of nonadiabatic transitions. 

Another challenge we experienced was the adequate description of the initial conditions for the (NA+BO)MD simulations. Since the ring-puckering mode exhibits a double-well potential and is highly anharmonic, the harmonic approximation underlying the usual (harmonic) Wigner sampling may lead to artifacts (as observed recently for another anharmonic molecule\cite{doi:10.1021/acs.jpca.3c02333}). More involved approaches to sampling the ground-state density are necessary, like \textit{ab initio} Born--Oppenheimer molecular dynamics coupled to a quantum thermostat used in the present work. The effects of the laser pulse are also nontrivial to include when dealing with a narrow spectral bandwidth of about 0.05 eV. The usual vertical approximation, even with an energy windowing, remains a crude approximation to the real excitation process and may lead to a photoexcited molecule with higher internal energy -- problematic when the photochemical processes involve a barrier in the excited electronic state. We believe that further methodological work on sampling the ground-state density, including the effect of a laser pulse explicitly, and electronic-structure theory are necessary to address these challenges.

While uncomfortable, the time constraint imposed by this prediction challenge to perform our simulations is actually a common scenario when collaborating with experimentalists, who may require rapid theoretical support to analyze their data. Thus, this situation overall calls for the development of diagnostic tools that would \textit{at least} allow us to test the sensitivity of fast predictions for more reliable interpretations of experimental results. 

\section*{Supplementary Material}

The Supplementary Material contains a depiction of the active-space orbitals, a detailed electronic-structure benchmark for the ground- and excited-electronic states, an analysis of the higher excited electronic states of cyclobutanone and their potential influence on the nonadiabatic dynamics, the detailed classification of photoproducts for the analysis of the nonadiabatic molecular dynamics, an analysis of the influence of spin-orbit coupling and intersystem crossing, a benchmark of the electronic structure for the ground-state of C$_3$ products, a correlation between interatomic distances and UED signal for the LIIC pathways, a comparison between the UED PDF signal for a large number of ground-state geometries and the subset of these geometries used as initial conditions for the dynamics, an analysis of the hopping geometries using multidimensional scaling, details about the fitting procedure for the populations, and a comparison between LZSH and FSSH dynamics. A zip file is provided with all the critical geometries located (xyz format) as well as the time-independent basis functions for the $\Delta$PDF and $\Delta I/I$ signal of each (ground-state) photoproduct (represented in Fig.~\ref{fig:dpdf_decomp}). 

\begin{acknowledgments}
The authors would like to thank CECAM for supporting the organization of the CECAM workshop 'Triggering out-of-equilibrium dynamics in molecular systems' in March 2023, which led to the idea of this prediction challenge. This project has received funding from the European Research Council (ERC) under the European Union's Horizon 2020 research and innovation programme (Grant agreement No. 803718, project SINDAM) and the EPSRC Grants EP/V026690/1. J.J. and P.S. were supported by the Czech Science Foundation, project number 23-07066S. This work was supported by the grant of Specific university research – grant No A2\_FCHI\_2023\_048. This article is based upon work from COST Action CA18212 - Molecular Dynamics in the GAS phase (MD-GAS), supported by COST (European Cooperation in Science and Technology).
\end{acknowledgments}

\section*{Data Availability Statement}

The data that support the findings of this study are available within the article and its supplementary material.


%

\end{document}


\newcommand\bfec[1]{\textcolor{orange}{\textit{ XXX BFEC: #1 XXX }}}
\newcommand\jj[1]{\textcolor{cyan}{\textbf{ JJ: #1}}}

\setlength{\tabcolsep}{12pt}

\DeclareFontFamily{OT1}{cmbr}{\hyphenchar\font45 }
\DeclareFontShape{OT1}{cmbr}{m}{n}{%
  <-9>cmbr8
  <9-10>cmbr9
  <10-17>cmbr10
  <17->cmbr17
}{}
\DeclareFontShape{OT1}{cmbr}{m}{sl}{%
  <-9>cmbrsl8
  <9-10>cmbrsl9
  <10-17>cmbrsl10
  <17->cmbrsl17
}{}
\DeclareFontShape{OT1}{cmbr}{m}{it}{%
  <->ssub*cmbr/m/sl
}{}
\DeclareFontShape{OT1}{cmbr}{b}{n}{%
  <->ssub*cmbr/bx/n
}{}
\DeclareFontShape{OT1}{cmbr}{bx}{n}{%
  <->cmbrbx10
}{}

\renewcommand{\thefigure}{S\arabic{figure}}
\renewcommand{\thetable}{S\arabic{table}}

\renewcommand{\rmdefault}{cmbr}
\renewcommand{\sfdefault}{cmbr}

\title{Supplementary Material: \\ Predicting the photodynamics of cyclobutanone triggered by a laser pulse at 200 nm and its MeV-UED signals -- a trajectory surface hopping and XMS-CASPT2 perspective}
\author{Ji\v{r}\'{i} Jano\v{s}}%
\affiliation{Department of Physical Chemistry, University of Chemistry and Technology, Technická 5, Prague 6, 166 28, Czech Republic}%
\affiliation{Centre for Computational Chemistry, School of Chemistry, University of Bristol, Bristol BS8 1TS, United Kingdom}%

\author{Joao Pedro Figueira Nunes}%
\affiliation{Diamond Light Source Ltd, Didcot, UK.}%

\author{Daniel Hollas}
\affiliation{Centre for Computational Chemistry, School of Chemistry, University of Bristol, Bristol BS8 1TS, United Kingdom}%

\author{Petr Slavíček}%
\email{petr.slavicek@vscht.cz}
\affiliation{Department of Physical Chemistry, University of Chemistry and Technology, Technická 5, Prague 6, 166 28, Czech Republic}%

\author{Basile F. E. Curchod}
\email{basile.curchod@bristol.ac.uk}
\affiliation{Centre for Computational Chemistry, School of Chemistry, University of Bristol, Bristol BS8 1TS, United Kingdom}%

\date{\today}

\maketitle

\renewcommand{\baselinestretch}{1.2}\normalsize
\tableofcontents   

\renewcommand{\baselinestretch}{1.2}\normalsize

\clearpage
\section{Active-space orbitals}

\begin{figure}[ht]
    \centering
    \includegraphics[width=0.8\textwidth]{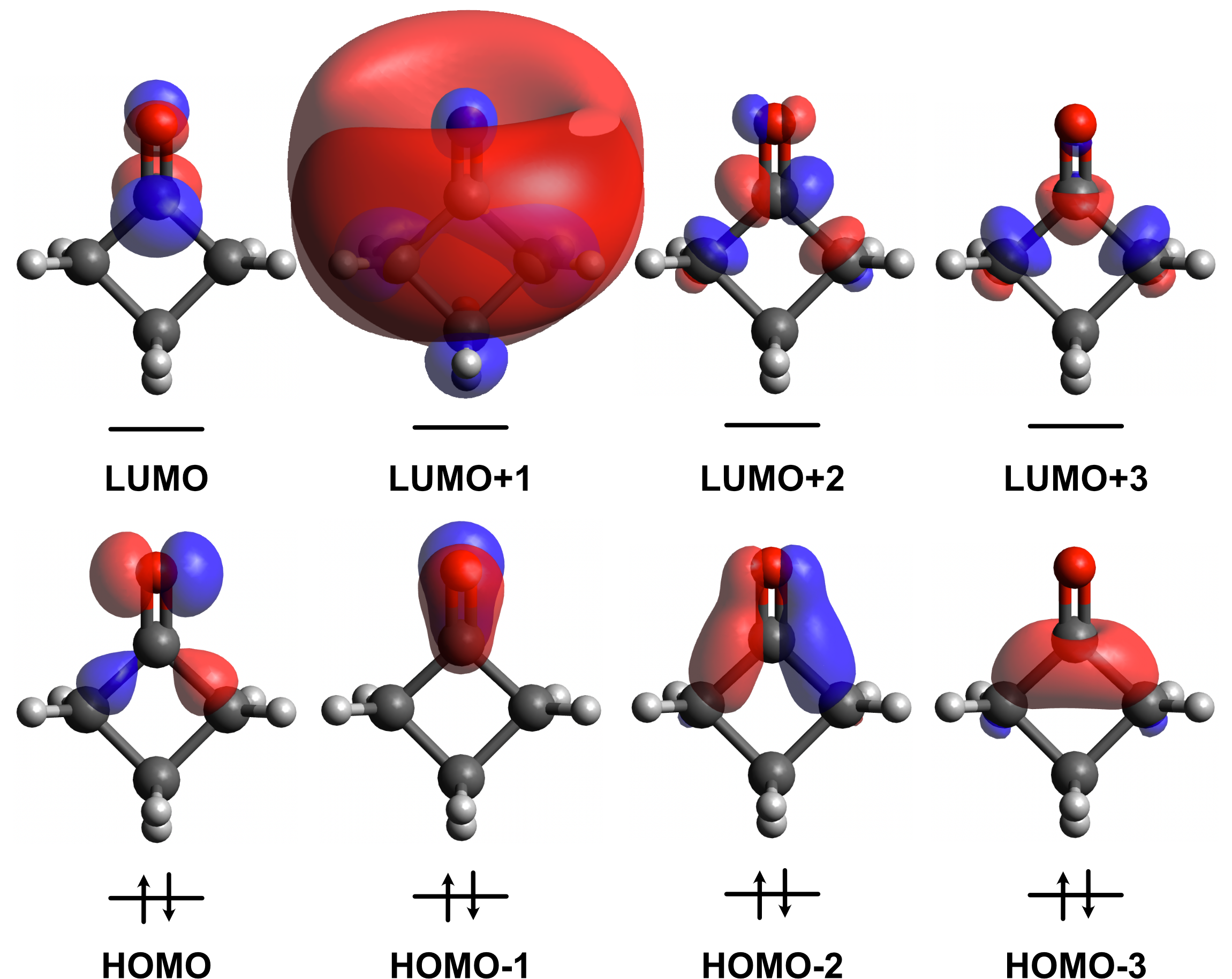}
    \caption{Active space orbitals employed in the XMS(3)-CASPT2(8/8)/aug-cc-pVDZ calculations, represented here for the ground-state optimized geometry obtained with XMS(3)-CASPT2(8/8)/aug-cc-pVDZ. Isovalue was set to 0.1 for all orbitals except the Rydberg one, for which an isovalue of 0.02 was used.}
    \label{fig:xms88_orbitals}
\end{figure}

\clearpage
\section{Electronic-structure benchmark for the ground- and excited-electronic states}

\subsection{Ground-state geometry of cyclobutanone}

To investigate the ground-state geometry of cyclobutanone, we first optimized the S$_0$ minimum-energy geometry with a series of exchange-correlation functionals and wavefunction-based methods implemented in Gaussian 09, Revision D.01\cite{g09}, see Table \ref{tab:gs_geometry}. The results show important variations for the ring-puckering dihedral angles between the carbon atoms, ranging from \ang{0.0} for BLYP to \ang{14.9} for MP2/cc-pVDZ and XMS(3S)-CASPT2(8/8)/aug-cc-pVDZ. While MP2, CCSD, and XMS-CASPT2 predict a ring-puckering vibrational wavenumber consistently around \SI{100}{cm^{-1}}, CASSCF and most DFT approaches estimate substantially lower value around \SI{30}{cm^{-1}}. One interpretation for this trend would be that a fine description of dynamic correlation is necessary to unravel the ring puckering (for example, the lack of dispersion interaction in standard exchange-correlation functionals and CASSCF results in a small ring puckering, \ang{0.0}-\ang{3.6}). Including a dispersion correction (B3LYP-D3 and \textomega B97XD) leads to an increase of both the ring puckering angle and its corresponding wavenumber, getting closer to the MP2/CCSD results. We decided to use \textomega B97XD functional for our DFT and LR-TDDFT/TDA calculations as it outperforms the other tested functionals.

\begin{table}[!ht]
    \centering
    \begin{tabular}{llcc}
\hline
method    & basis set   & $\theta_\mathrm{ring. puck.}$ (\textdegree) & $\tilde{\nu}_\mathrm{ring. puck.}$ (cm$^{-1}$) \\
\hline
BLYP                    & aug-cc-pVTZ                 & 0.0  & 32.6  \\
B3LYP                   & aug-cc-pVTZ                 & 3.8  &  6.7  \\ 
B3LYP-D3                & aug-cc-pVTZ                 & 4.3  & 12.8  \\ 
CAM-B3LYP               & aug-cc-pVTZ                 & 2.3  & 18.7  \\
\textomega B97XD        & aug-cc-pVTZ                 & 4.7  & 38.3 \\
\textomega B97XD        & aug-cc-pVDZ                 & 7.5  & 66.5 \\
\hline
MP2                     & cc-pVDZ                     & 14.9 & 117.2 \\
MP2                     & cc-pVTZ                     & 14.3 & 108.0 \\
MP2                     & aug-cc-pVDZ                 & 14.6 & 113.6 \\
MP2                     & aug-cc-pVTZ                 & 13.9 & 104.4 \\
\hline
CCSD                    & cc-pVDZ                     & 11.8 & 95.9  \\
CCSD                    & aug-cc-pVDZ                 & 11.7 & 98.3  \\
\hline
SA(3S,3T)-CASSCF(8/8)  & 6-31+G*                     & 3.6  & 38.0  \\
\hline
XMS(3S)-CASPT2(8/8) & aug-cc-pVDZ      &  14.9 & 111.8  \\
\hline
    \end{tabular}
    \caption{Ring-puckering dihedral angle ($\theta_\mathrm{ring. puck.}$, defined by the four carbon atoms) and its harmonic vibrational wavenumber ($\tilde{\nu}_\mathrm{ring. puck.}$) at the ground-state minimum of cyclobutanone obtained with the corresponding electronic structure method.}
    \label{tab:gs_geometry}
\end{table}

To further investigate the anharmonicity of the ring-puckering mode, we optimized the transition state characterizing the ring inversion of cyclobutanone with a selection of electronic-structure methods (Table~\ref{tab:ts_puckering}). We could not localize a transition state for BLYP as this functional does not show a double well potential and predicts a planar ground-state minimum geometry and also for B3LYP. The CAM-B3LYP functional shows a very shallow double-well potential, with a tiny activation barrier for the ring inversion. The \textomega B97XD functional (which includes a dispersion correction) reproduces a well-defined minimum and transition state and, combined with the aug-cc-pVDZ basis set, predicts an activation barrier comparable to the experimentally measured value. The MP2 method also finds a well-defined minimum and transition state but with a barrier significantly higher than the experimental value.\cite{Scharpen1968} We should emphasize, though, that the value is within a chemical accuracy of 1 kcal/mol. Thus, we selected the MP2 and \textomega B97XD methods for our investigation of the ground electronic state of cyclobutanone as they both predict the double-well potential suggested experimentally.\cite{Scharpen1968}

\begin{table}[ht]
\begin{tabular}{llccc}
\hline
method    & basis set   & $\tilde{\nu}_\mathrm{ring. puck.}$ (cm$^{-1}$) & $\tilde{\nu}_\mathrm{TS}$ (cm$^{-1}$) & $\Delta E^\ddagger$ (kcal/mol) \\
\hline
CAM-B3LYP & aug-cc-pVTZ & 18.7              & $15.8$i                  & 0.0003   \\
\textomega B97XD    & aug-cc-pVTZ & 38.3              & $38.7$i                  & 0.0068   \\
\textomega B97XD    & aug-cc-pVDZ & 66.5              & $49.1$i                  & 0.0351    \\
MP2       & cc-pVDZ     & 117.2             & $91.3$i                  & 0.4079    \\
MP2       & aug-cc-pVDZ & 113.6             & $98.6$i                  & 0.4336    \\
CCSD(T)   & aug-cc-pVDZ & 65.0             & $144.0$i                  & 0.1957    \\
\hline   
experiment\cite{Scharpen1968} & & - & - & 0.0217 \\
\hline   
\end{tabular}
\caption{Ring-puckering harmonic vibrational wavenumber ($\tilde{\nu}_\mathrm{ring. puck.}$) at the ground-state minimum, imaginary vibrational wavenumber ($\tilde{\nu}_\mathrm{TS}$) at the transition state, and activation barrier ($\Delta E^\ddagger$) for the ring inversion of cyclobutanone. The minima and transition states were optimized with the corresponding electronic-structure method except for the CCSD(T) calculation which used the MP2/aug-cc-pVDZ geometries.}
\label{tab:ts_puckering}
\end{table}

\clearpage
\subsection{Benchmark of the excitation energies in the Franck-Condon region}

In this Section, we summarize the benchmark of electronic-structure methods we conducted to test the excitation energies and oscillator strengths. Two different coupled-clusters methods (EOM-CC3 and EOM-CCSD) and two different exchange-correlation functionals (CAM-B3LYP and \textomega B97XD) for LR-TDDFT/TDA were used, combined with two basis sets aug-cc-pVDZ (Table~\ref{tab:se_exc_1}) and aug-cc-pVTZ (Table~\ref{tab:se_exc_2}). No significant differences in excitation energies and oscillator strengths can be observed between the different strategies tested. 

\begingroup
\squeezetable
\begin{table}[ht]
\begin{tabular}{ccccccccc}
\hline
\multirow{2}{*}{state} & \multicolumn{2}{c}{EOM-CC3}        & \multicolumn{2}{c}{EOM-CCSD}   & \multicolumn{2}{c}{\makecell{LR-TDDFT/TDA\\CAM-B3LYP}} & \multicolumn{2}{c}{\makecell{LR-TDDFT/TDA\\\textomega B97XD}} \\
                       & $\Delta E$ (eV) & \textit{$f$} & $\Delta E$ (eV) & \textit{$f$} & $\Delta E$ (eV)   & \textit{$f$}  & $\Delta E$ (eV) & \textit{$f$} \\
\hline
S$_1$ & 4.39 & 0.0000 & 4.41 & 0.0000 & 4.37 & 0.0000 & 4.38 & 0.0000 \\
S$_2$ & 6.36 & 0.0380 & 6.43 & 0.0378 & 6.42 & 0.0401 & 6.58 & 0.0376 \\
S$_3$ & 6.96 & 0.0016 & 7.05 & 0.0022 & 7.02 & 0.0015 & 7.15 & 0.0019 \\
S$_4$ & 7.12 & 0.0000 & 7.19 & 0.0001 & 7.17 & 0.0000 & 7.33 & 0.0001 \\
S$_5$ & 7.15 & 0.0021 & 7.23 & 0.0016 & 7.20 & 0.0005 & 7.33 & 0.0003 \\
\hline
T$_1$ & -    & -      & 4.05 & - & 3.81 & - & 3.85 & - \\
T$_2$ & -    & -      & 6.31 & - & 6.07 & - & 6.16 & - \\
T$_3$ & -    & -      & 6.35 & - & 6.33 & - & 6.49 & - \\
T$_4$ & -    & -      & 6.99 & - & 6.97 & - & 7.09 & - \\
T$_5$ & -    & -      & 7.12 & - & 7.08 & - & 7.18 & - \\
\hline
\end{tabular}
\caption{Excitation energies ($\Delta E$) and oscillator strengths ($f$) obtained at different levels of electronic-structure theory on the support of the ground-state geometry of cyclobutanone (MP2/cc-pVTZ). All excited-state methods employed the aug-cc-pVDZ basis set.}
\label{tab:se_exc_1}
\end{table}
\endgroup

\begingroup
\squeezetable
\begin{table}[ht]
\begin{tabular}{ccccccccc}
\hline

\multirow{2}{*}{state} & \multicolumn{2}{c}{EOM-CC3}        & \multicolumn{2}{c}{EOM-CCSD}   & \multicolumn{2}{c}{\makecell{LR-TDDFT/TDA\\CAM-B3LYP}} & \multicolumn{2}{c}{\makecell{LR-TDDFT/TDA\\\textomega B97XD}} \\
                  & $\Delta E$ (eV) & \textit{$f$} & $\Delta E$ (eV) & \textit{$f$} & $\Delta E$ (eV)   & \textit{$f$}  & $\Delta E$ (eV) & \textit{$f$} \\
\hline
S$_1$ & 4.38 & 0.0000 & 4.43 & 0.0000 & 4.39 & 0.0000 & 4.40 & 0.0000 \\
S$_2$ & 6.48 & 0.0361 & 6.61 & 0.0361 & 6.44 & 0.0385 & 6.57 & 0.0357 \\
S$_3$ & 7.07 & 0.0016 & 7.22 & 0.0023 & 7.02 & 0.0014 & 7.11 & 0.0015 \\
S$_4$ & 7.21 & 0.0000 & 7.34 & 0.0001 & 7.16 & 0.0000 & 7.27 & 0.0005 \\
S$_5$ & 7.25 & 0.0014 & 7.40 & 0.0009 & 7.19 & 0.0004 & 7.27 & 0.0000 \\
\hline
T$_1$ & -    & -      & 4.08 & - & 3.84 & - & 3.87 & - \\
T$_2$ & -    & -      & 6.28 & - & 6.07 & - & 6.14 & - \\
T$_3$ & -    & -      & 6.53 & - & 6.35 & - & 6.47 & - \\
T$_4$ & -    & -      & 7.16 & - & 6.97 & - & 7.05 & - \\
T$_5$ & -    & -      & 7.28 & - & 7.08 & - & 7.12 & - \\
\hline
\end{tabular}
\caption{Excitation energies ($\Delta E$) and oscillator strengths ($f$) obtained at different levels of electronic-structure theory on the support of the ground-state geometry of cyclobutanone (MP2/cc-pVTZ). All excited-state methods employed the aug-cc-pVTZ basis set.}
\label{tab:se_exc_2}
\end{table}
\endgroup

\clearpage
\subsection{Electronic energies diagram}

\begin{figure}[ht!]
    \centering
    \includegraphics[width=\textwidth]{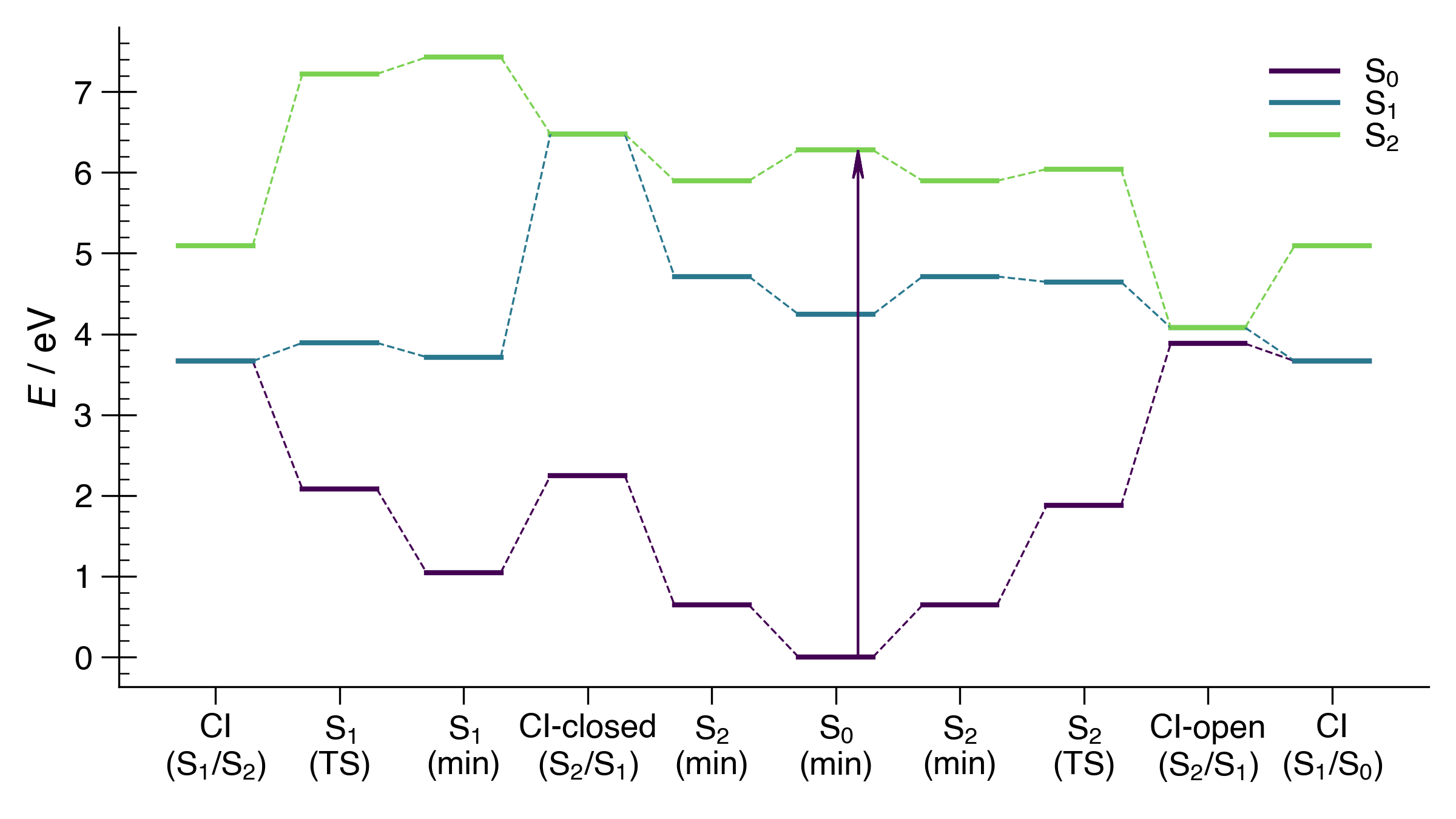}
    \caption{Electronic energies for the geometries used to generate the LIICs together with transition states in S$_2$ and S$_1$ calculated at the XMS-CAPT2(8/8)/aug-cc-pVDZ level.}
    \label{fig:pes_mapping}
\end{figure}

\clearpage
\subsection{Benchmark of the electronic-structure methods along the LIIC pathways}

\begin{figure}[ht!]
    \centering
    \includegraphics[width=\textwidth]{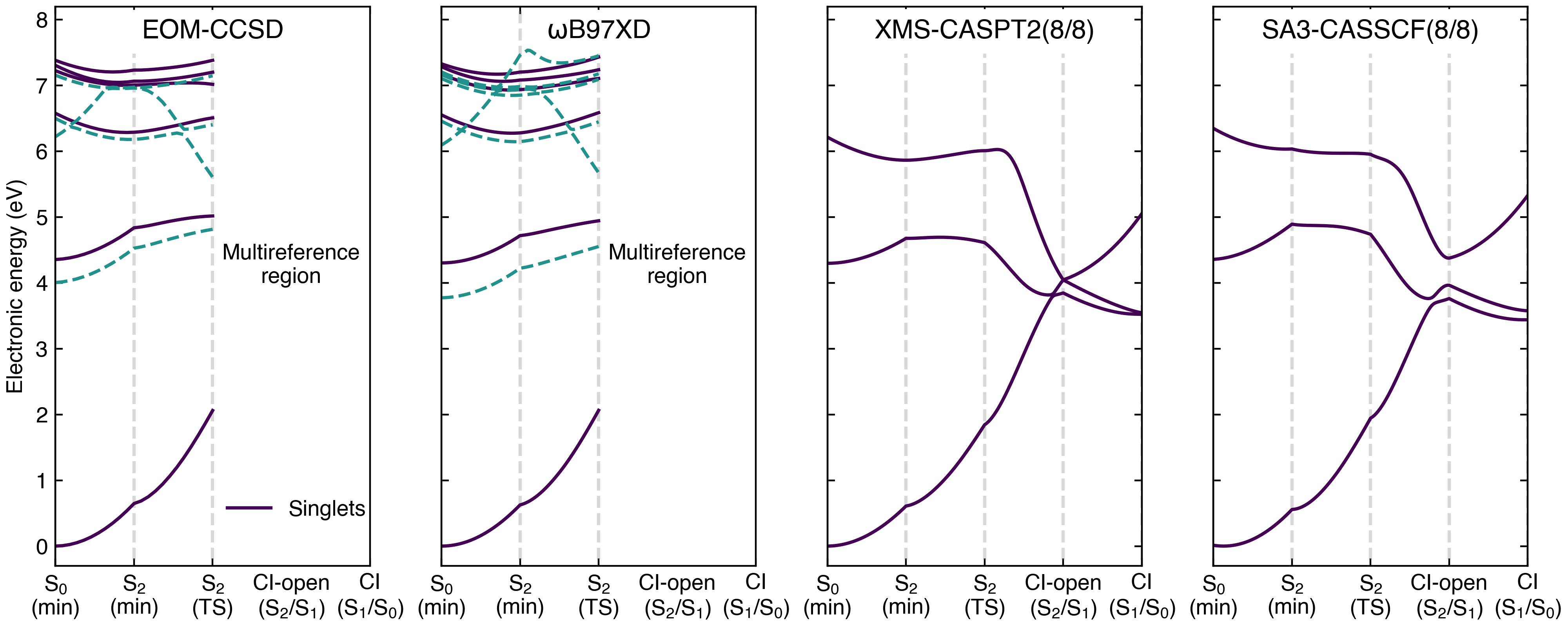}
    \caption{Electronic energies along the LIIC for the adiabatic pathway obtained with EOM-CCSD/aug-cc-pVTZ, LR-TDDFT/TDA/\textomega B97XD/aug-cc-pVDZ, XMS-CASPT2(8/8)/aug-cc-pVDZ and SA(3S,3T)-CASSCF/6-31+G*. We plot S$_0$-S$_2$ and T$_1$-T$_5$ where available.}
    \label{fig:liic_bench}
\end{figure}

We present in Fig.~\ref{fig:liic_bench} a benchmark of several electronic structure methods --  EOM-CCSD/aug-cc-pVTZ, LR-TDDFT/TDA/\textomega B97XD/aug-cc-pVDZ, XMS-CASPT2(8/8)/aug-cc-pVDZ, and SA3-CASSCF(8/8)/aug-cc-pVDZ --  using the LIIC for the adiabatic pathway. The LIIC discussed here also includes the geometry of the S$_2$ transition state geometry, S$_2$(TS), between S$_2$(min) and CI-open(S$_2$/S$_1$). Once cyclobutanone ring starts to open around the S$_2$(TS) geometry, a multiconfigurational wavefunction is required, and both EOM-CCSD and \textomega B97XD begin to fail -- we therefore only report their results up to the S$_2$(TS) geometry. All the methods show a comparable behavior in the region between S$_0$(min) and S$_2$(TS) -- explaining the good performance of LR-TDDFT/TDA/\textomega B97XD in reproducing the vibronic structure of the S$_2$$\leftarrow$S$_0$ band in the photoabsorption cross-section of cyclobutanone. The lack of dynamic correlation in SA3-CASSCF causes almost  barrierless ring opening in the S$_2$ state. EOM-CCSD and LR-TDDFT/TDA/\textomega B97XD show that the S$_3$-S$_5$ (Rydberg $3p$) states evolve in parallel to the S$_2$ electronic state. Comparing energies of the triplet states, we again observe an agreement between applied approaches. The characters of the triplet states in the S$_0$ minimum-energy geometry are following: T$_1$($n\pi^*$), T$_2$($\pi\pi^*$), T$_3$($3s$), T$_4$($3p$), and T$_5$($3p$).

\clearpage
\section{Higher excited electronic states of cyclobutanone and their possible influence on the nonadiabatic dynamics}

\begin{figure}[ht!]
    \centering
    \includegraphics[width=\textwidth]{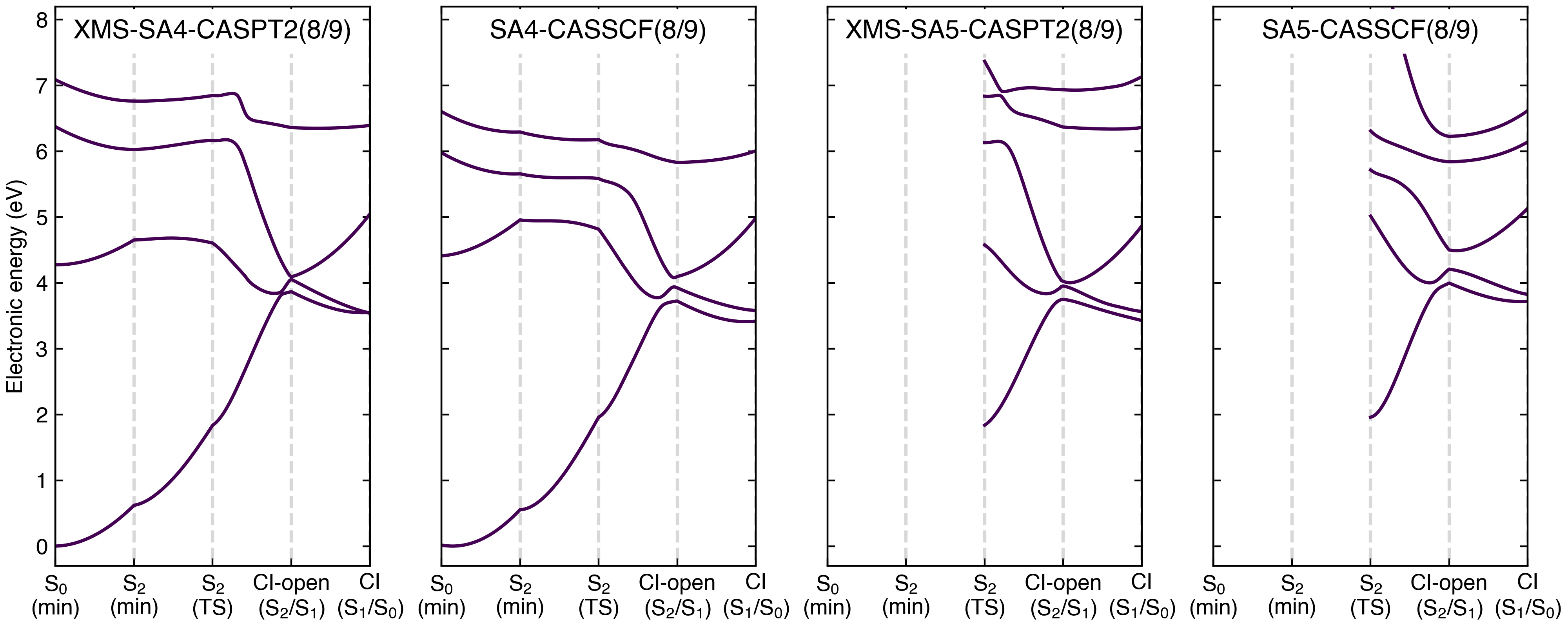}
    \caption{Electronic energies along the adiabatic LIIC pathway obtained with XMS-SA4-CASPT2(8/9), SA4-CASSCF(8/9) XMS-SA5-CASPT2(8/9), and SA4-CASSCF(8/9) using the aug-cc-pVDZ basis set.}
    \label{fig:multi_liic}
\end{figure}

To rule out possible adiabatic trapping when the molecule crosses the ring-opening transition state in the S$_2$ state, we have performed also XMS-CASPT2 and CASSCF calculations considering 4 and 5 states in the state-averaging procedure along the adiabatic pathway LIIC, see Fig.~\ref{fig:multi_liic}. In order to capture the 3p Rydberg state, we extended our active space with one 3p Rydberg orbital. As a consequence, we will have only one Rydberg state in our calculations. However, the previous EOM-CCSD and \textomega B97XD calculations have shown that they evolve in parallel. In the case of 5 states in the state averaging, the other state was a $\sigma^*$ state accounting for dissociation. Unfortunately, this calculation was stable only between S$_2$ (TS) and CI (S$_1$/S$_0$). The calculations show that the S$_3$ excited state is well separated from the S$_2$ electronic state and a possible diabatic trapping is very unlikely. Thus, not including higher excited states allowing for upward transitions should be reasonable.

\clearpage
\section{Classification tree for the analysis of the nonadiabatic molecular dynamics}

\begin{figure}[ht]
    \centering
    \includegraphics[width=0.8\textwidth]{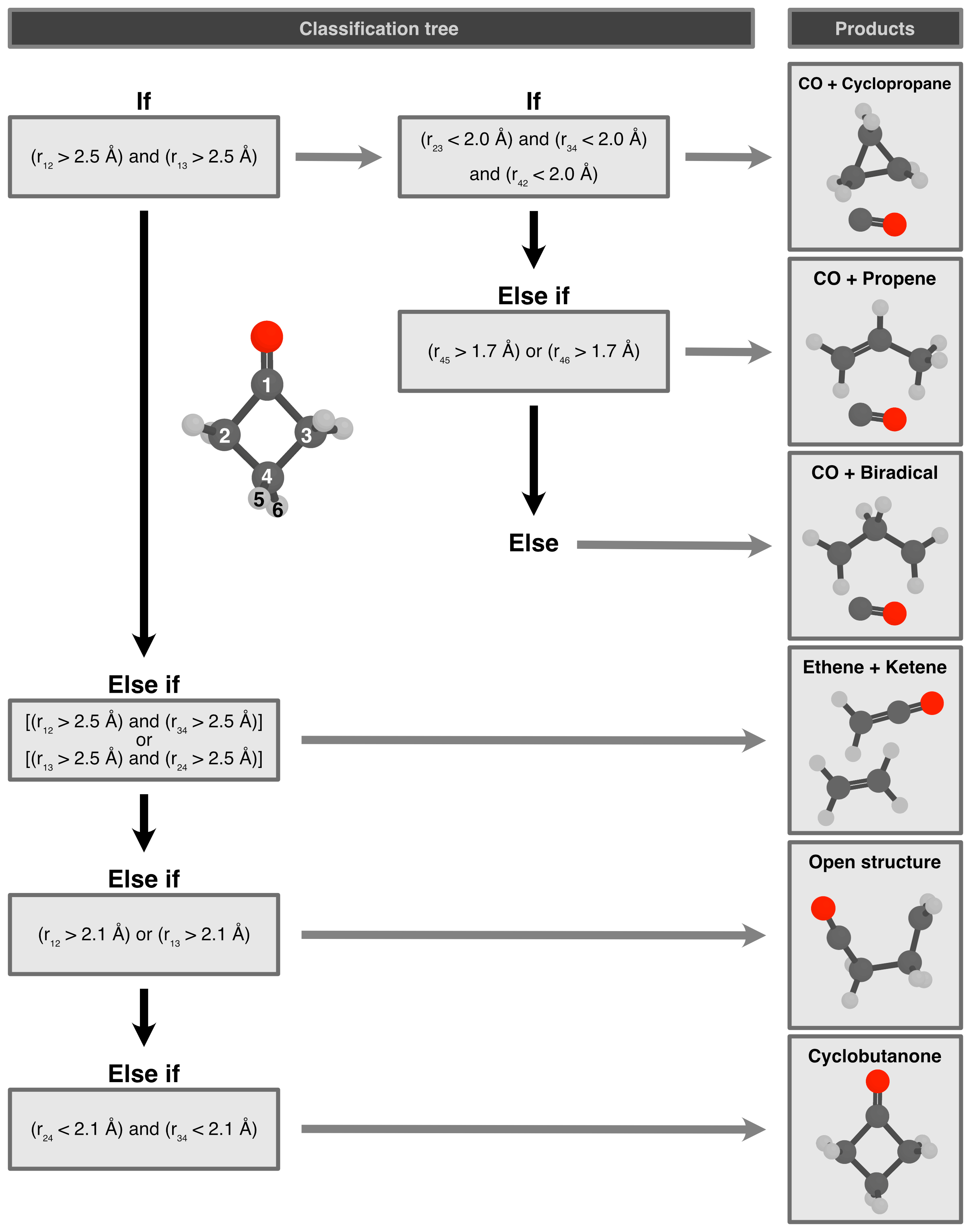}
    \caption{Classification tree used to extract the time-dependent products from the nonadiabatic molecular dynamics. The structures are classified into six categories, which were identified based on a visual inspection of the trajectories. Black arrows represent the 'false' branch of the condition, while grey ones identify the 'true' branch.}
    \label{fig:si_classification}
\end{figure}

\clearpage
\section{Spin-orbit coupling and intersystem crossing}

\subsection{Spin-orbit coupling along the LIIC for adiabatic pathway}

\begin{figure}[ht!]
    \centering
    \includegraphics[width=1.0\textwidth]{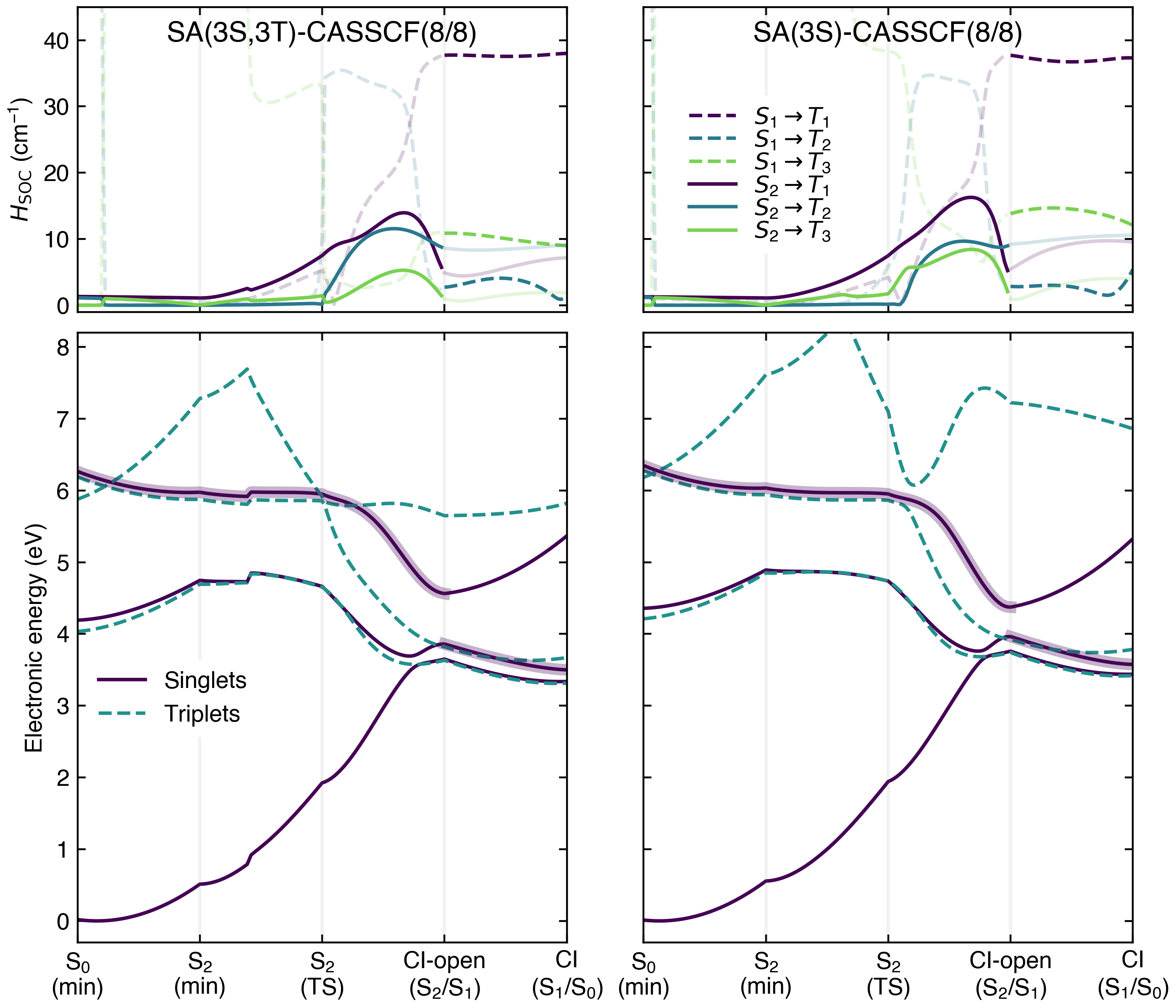}
    \caption{The SOC magnitude calculated along the LIIC of the adiabatic path (including also the transition state) with SA(3S,3T)-CASSCF(8/8)/aug-cc-pVDZ and SA(3S)-CASSCF(8/8)/aug-cc-pVDZ methods The active state likely to drive the dynamics is highlighted together with the SOC between the active space and the remaining triplets.}
    \label{fig:soc}
\end{figure}

To estimate the SOC along the adiabatic pathway, we performed CASSCF(8/8)/aug-cc-pVDZ calculations along the LIIC in the Molpro2012 package.\cite{molpro2012} The active space is the same as described in the main article. The characters of the triplet states in the S$_0$ minimum are $n\pi^*$, $\pi\pi^*$ and $3s$. We adopted two strategies for the state averaging: (\textit{i}) we averaged over all singlet and triplet states [denoted as SA(3S,3T)], and (\textit{ii}) we averaged only over the singlet states and the triplet states were calculated in the CASCI manner [denoted as SA(3S)]. While the strategy (\textit{i}) yields more reliable triplet states, it exhibits a root flipping between S$_2$ (min) and S$_2$ (TS) which causes a discontinuity in energies. The root flipping happens for the $\pi\pi^*$ changing with $n\sigma^*$ state. The strategy (\textit{ii}) avoids this by using only singlets for orbital optimization and, therefore, provides continuous energy curves. Note that both approaches yield comparable results with similar values of SOC. The SOC was calculated using the Pauli-Breit Hamiltonian as described in the main article. The results are depicted in Figure~\ref{fig:soc}. The SOC magnitude between the active state likely to drive the dynamics and the neighbouring triplet states is below 20~cm$^{-1}$ along the whole LIIC coordinate except between the S$_1$ and T$_1$ states for the open structure where the SOC is 40~cm$^{-1}$. Such small values combined with short lifetimes of cyclobutanone in the S$_2$ and S$_1$ states suggest that ISC is unlikely.

\subsection{LZSH dynamics with intersystem crossing}

To further test our hypothesis that intersystem crossing will not compete with internal conversion, we performed LZSH dynamics including triplet states\cite{Suchan2020} implemented in the ABIN code.\cite{Hollas2019} The electronic quantities were calculated at the SA(3S,3T)-CASSCF(8/8)/6-31+g* level in Molpro2012 package.\cite{molpro2012} We launched only 10 initial conditions for this purpose. The LZSH extension for triplet states evaluates SOC every time any triplet state crosses the propagated singlet and calculates hopping probability. We note that the CASSCF/LZSH dynamics showed qualitatively similar behaviour as XMS-CASPT2/FSSH simulations observing the same products only with substantially shorter lifetimes. We evaluated the SOCs between the current singlet state and all triplets, see Table~\ref{tab:soc_lzsh}. The values of SOC when the trajectories were in the S$_2$ were negligible with only one exception of 22.7~cm$^{-1}$ encountered at the top of the ring-opening barrier. Thus, we conclude that the ISC from S$_2$ to S$_1$ minima cannot compete with internal conversion. The couplings increased to an average value of 26~cm$^{-1}$ in the S$_1$ state yet this was always in the open biradical structure which was very quickly funnelled to S$_0$.

\setcounter{table}{4}
\begin{table}[!ht]
    \centering
    \begin{tabular}{c c c}
    \hline
        active state & max $H_\mathrm{SOC}$ (cm$^{-1}$) & average $H_\mathrm{SOC}$ (cm$^{-1}$)\\
        \hline
        S$_2$ & 22.7 & $2.3\pm4.1$ \\
        S$_1$ & 36.5 & $25.8\pm5.3$ \\
        S$_0$ & 27.8 & $7.1\pm9.7$ \\
        \hline
    \end{tabular}
    \caption{The SOC magnitudes between the active state driving the LZSH dynamics and the triplets electronic states crossing this driving state during the propagation.}
    \label{tab:soc_lzsh}
\end{table}

\clearpage
\section{Benchmarking the electronic energies relevant for the ground-state dynamics of C$_3$ products}

\begin{figure}[ht!]
    \centering
    \includegraphics[width=0.8\textwidth]{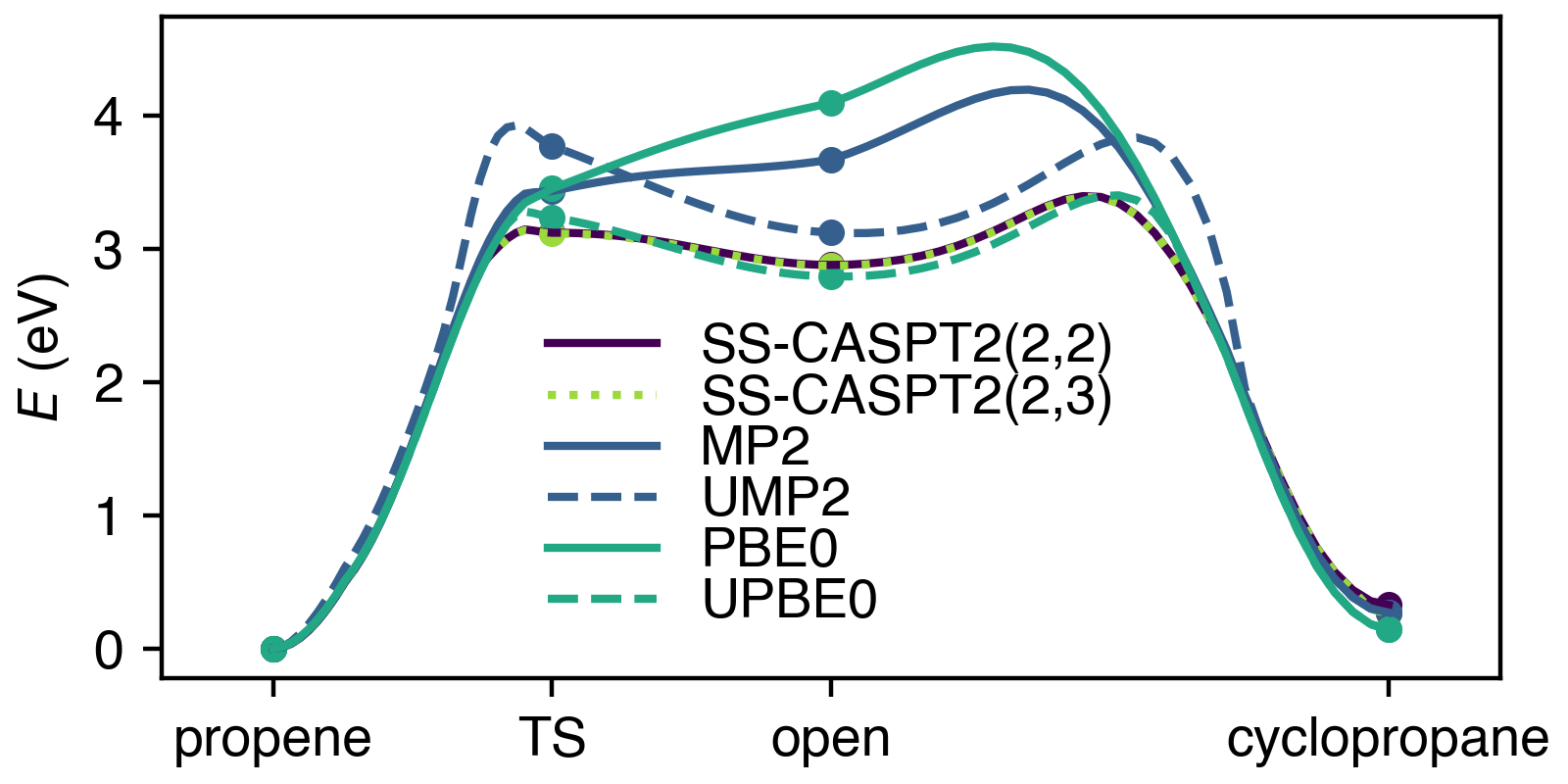}
    \caption{Comparison of electronic energies obtained from MP2, unrestricted MP2 (UMP2), PBE0 and unrestricted PBE0 (UPBE0) and single-state (SS) CASPT2 methods on the LIICs between propene, proton-transfer transition state (TS), open biradical structure and cyclopropane. The Figure shows that both propene and cyclopropane are stable products, while the open biradical structure is highly unstable. The Figure also validates the choice of electronic structure for the ground-state dynamics.}
    \label{fig:c3_pes}
\end{figure}

\clearpage
\section{Evolution of distances between atoms along LIICs for analysis of the UED signal.}

\begin{figure}[ht]
    \centering
    \includegraphics[width=1.0\textwidth]{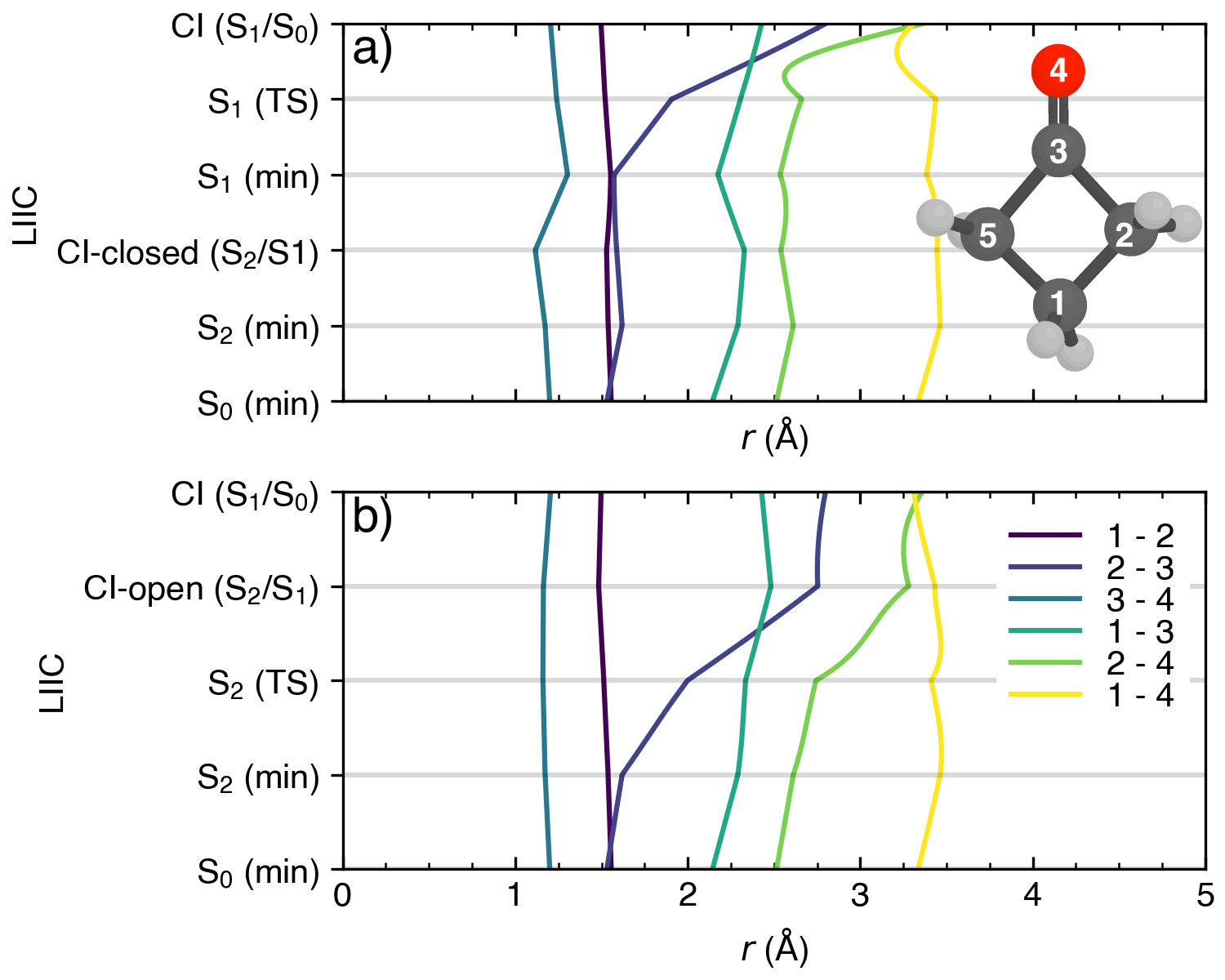}
    \caption{Selected distances between carbon and oxygen atoms in cyclobutanone along the LIICs for a) the CI pathway and b) the adiabatic pathway used for calculations of the UED signal in Figure~6.}
    \label{fig:liic_distances}
\end{figure}

\clearpage
\section{UED signal for the complete ground-state sampling and our set of initial conditions}

\begin{figure}[ht]
    \centering
    \includegraphics[width=0.8\textwidth]{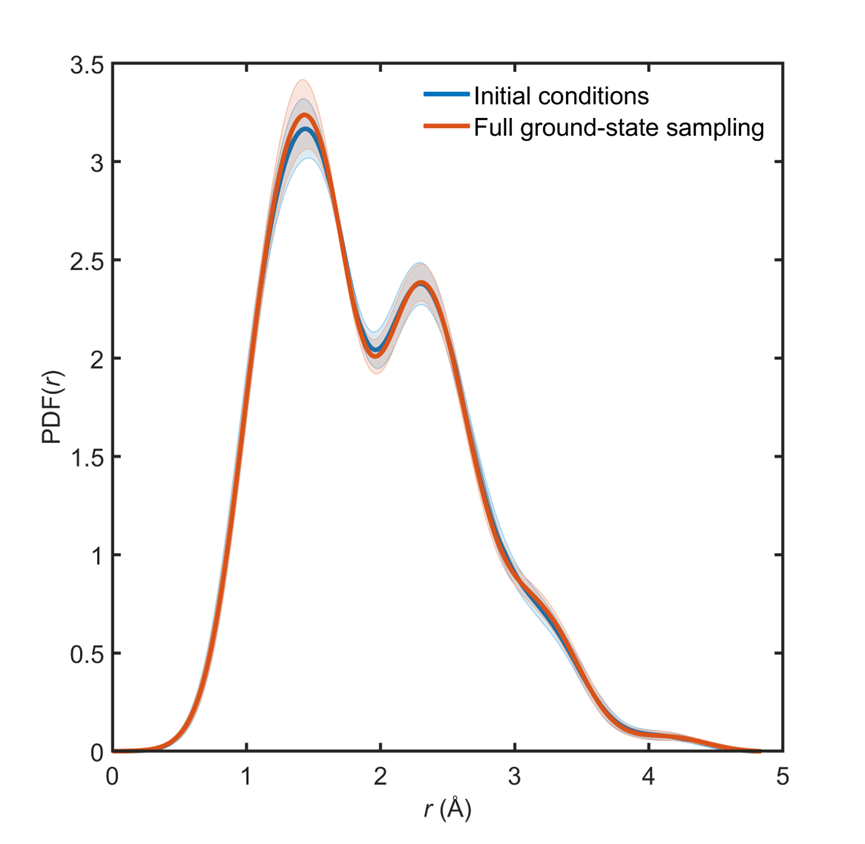}
    \caption{The steady-state PDF as an average of the PDF calculated for each initial condition used for the nonadiabatic dynamics (119 geometries) and the full ground-state sampling (5000 geometries) with a QT-BOMD at the MP2/cc-pVDZ level.}
    \label{fig:ued_ic_gs}
\end{figure}

\clearpage
\section{Classification of reaction pathways in FSSH trajectories with multidimensional scaling}

\begin{figure}[ht]
    \centering
    \includegraphics[width=1.0\textwidth]{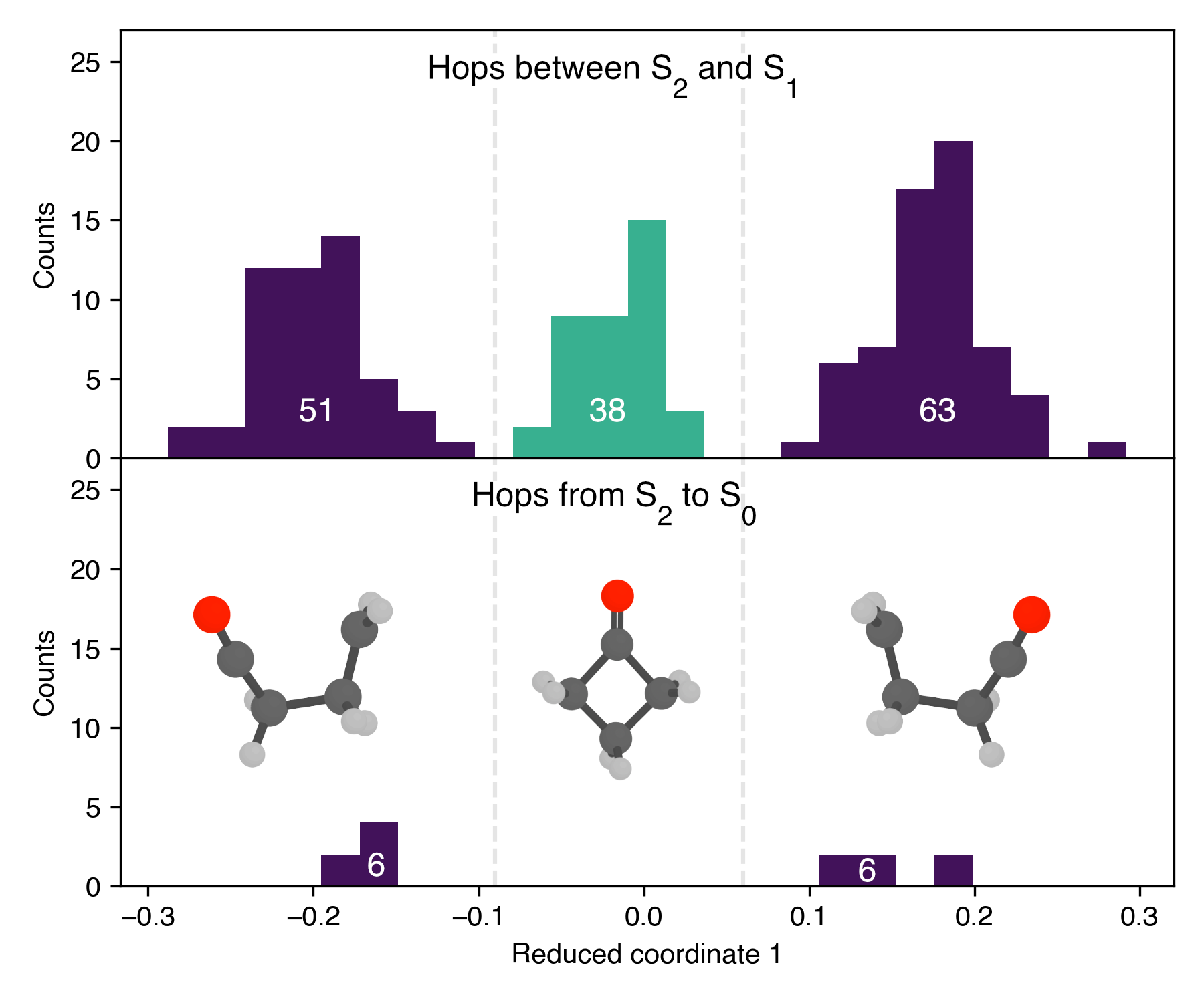}
    \caption{Representation of the hopping geometries, produced with MDS, between $\mathrm{S_2}$ and $\mathrm{S_1}$ states considering both up and down hops (upper panel) and hops between $\mathrm{S_2}$ and $\mathrm{S_0}$ states, where only down hops occurred (bottom panel). The purple color represents geometries resembling the CI-open (adiabatic pathway), while the green color highlights geometries resembling the CI-closed (CI pathway). The white numbers represent the total count of a given set of data.}
    \label{fig:mds}
\end{figure}

The reaction pathways were identified and classified by a multidimensional scaling (MDS) method following Ref.~\citenum{Janos2023}. MDS is a dimensionality reduction tool that can project a high-dimensional molecular geometry onto a low-dimensional space suitable for visual analysis. MDS reduces the dimensionality of a problem but tries to preserve distances between data points as much as possible. MDS is, therefore, suitable for comparing different clusters of data in reduced dimensionality. In the present work, we applied the MDS procedure to the hopping geometries observed between the S$_2$ and S$_1$ electronic states and from S$_2$ to S$_0$, which allowed us to highlight two possible reaction pathways. The two pathways differ by the shape of the conical intersection reached -- thus, the hopping geometries form two clusters. To calculate the distances, we first aligned the hopping geometries and then calculated the Cartesian distance between them (considering only carbon and oxygen atoms to reduce the noise coming from hydrogen atoms). Then, we ran the MDS procedure and plotted the data in the first reduced coordinate, where the clusters can easily be identified (see Fig.~\ref{fig:mds}). More information about the MDS method applied to molecular geometries can be found in Refs.~\citenum{Janos2023} and \citenum{Li2018}.

\clearpage
\section{Fitting of the electronic-state and relative photoproduct populations}

\subsubsection{Time-dependent electronic-state populations}

\begin{figure}[ht]
    \centering
    \includegraphics[width=1.0\textwidth]{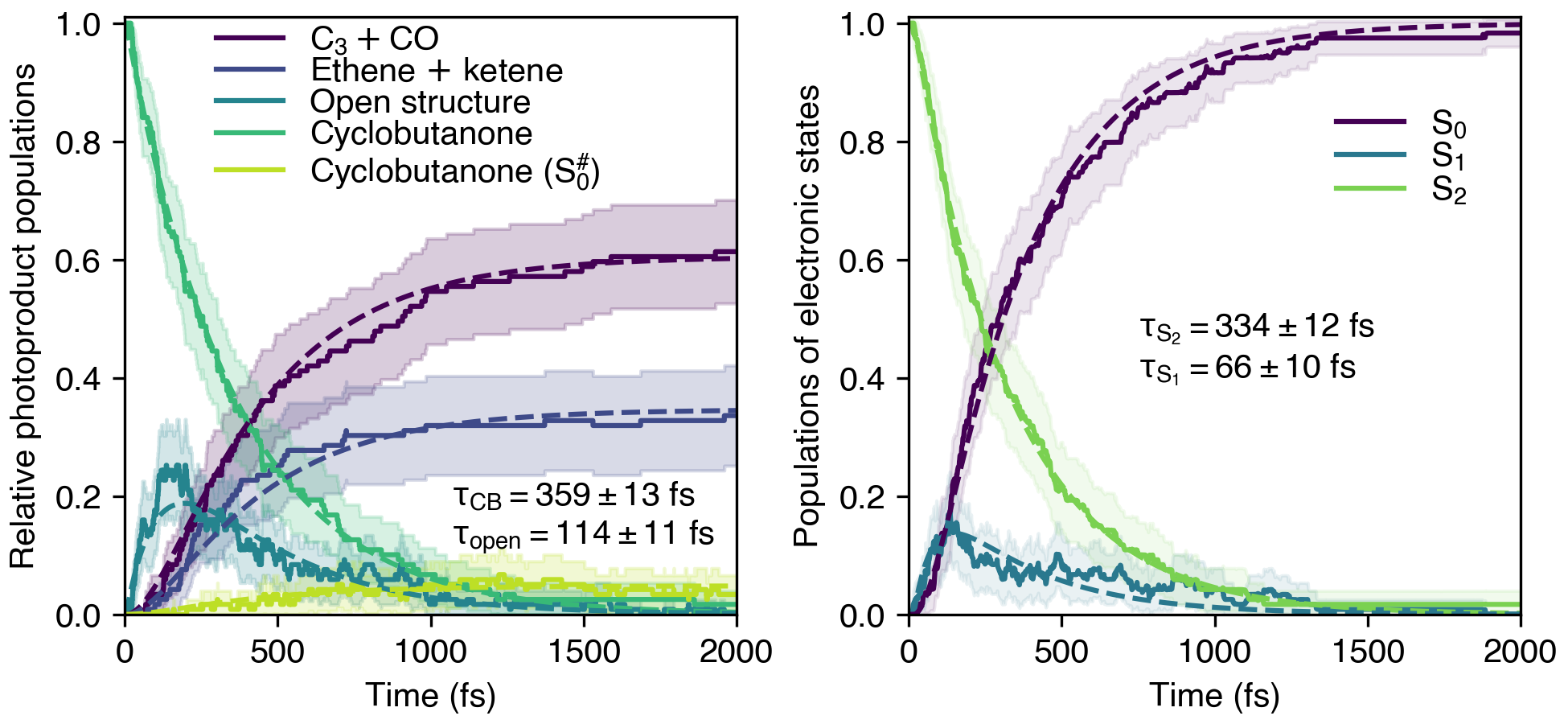}
    \caption{The time-dependent relative photoproduct populations (left) and electronic-state populations (right) from the FSSH/XMS-CASPT2 simulations.}
    \label{fig:fssh_fit}
\end{figure}

To fit the electronic-state population traces and obtain the lifetimes, we considered a consecutive reaction scheme $\mathrm{S}_2 \rightarrow \mathrm{S}_1 \rightarrow \mathrm{S}_0$ with an initial delay time $t_0$. The delay time corresponds to the initial period of the simulation when the nuclear wavepacket moves away from the Franck-Condon region, before the nonadiabatic transfer to the other electronic states begins. The equations for the time-dependent populations were derived from the kinetic scheme considering that the reactions start at time $t_0$. The S$_2$-state population trace was fitted to a delayed exponential function

\begin{equation}
    p_{\mathrm{S}_2}(t)=
    \begin{cases}
        1 & \text{if } t \in [0, t_0]\\
        \mathrm{e}^{-k_\mathrm{\mathrm{S}_2}(t-t_0)}  & \text{if } t > t_0
    \end{cases}
\end{equation}

\noindent where the delay time $t_0$ and the rate constant $k_\mathrm{\mathrm{S}_2}$ are the fitting parameters. The total lifetime of S$_2$ was evaluated as $\tau_{\mathrm{S}_2} = t_0 + 1/k_\mathrm{\mathrm{S}_2}$, i.e., the time it takes for the S$_2$ population to reach a value of $1/e$. The S$_1$ state appears as the intermediate in the kinetic scheme and the following function can be derived for its population

\begin{equation}
    p_{\mathrm{S}_1}(t)=
    \begin{cases}
        0 & \text{if } t \in [0, t_0]\\
        \frac{k_{\mathrm{S}_2}}{k_{\mathrm{S}_1}-k_{\mathrm{S}_2}} \cdot \left[ \mathrm{e}^{-k_\mathrm{\mathrm{S}_2}(t-t_0)} - \mathrm{e}^{-k_\mathrm{\mathrm{S}_1}(t-t_0)} \right] & \text{if } t > t_0
    \end{cases}
\end{equation}

\noindent where the rate constant $k_{\mathrm{S}_1}$ is the only fitting parameter. The $\mathrm{S}_1$ lifetime is then evaluated as $\tau_{\mathrm{S}_1}=1/k_\mathrm{S_1}$. The populations from the FSSH dynamics, the fit of the populations, and the lifetimes obtained are plotted in the right panel of Fig.~\ref{fig:fssh_fit}.

\subsubsection{Time-dependent relative photoproduct populations}

To fit the time-dependent relative photoproduct populations, we considered the following reaction scheme based on our FSSH simulations, consisting in three major channels:

\begin{align}
    \mathrm{Cyclobutanone} \xrightarrow[]{k_\mathrm{CB}} \mathrm{Open \; structure} &\xrightarrow[]{k_\mathrm{CO}} \mathrm{CO+C}_3 \\
    &\xrightarrow[]{k_\mathrm{EK}} \mathrm{Ethene+ketene} \\
    &\xrightarrow[]{k_\mathrm{CB^\#}} \mathrm{Cyclobutanone \; (S_0^\#)}
\end{align}

\noindent where the $k$s are the respective rate constants and $\mathrm{S_0^\#}$ signifies a hot ground-state cyclobutanone. The total rate constant for the decay of the open structure is given as

\begin{equation}
    k_\mathrm{open} = k_\mathrm{CO} + k_\mathrm{EK} + k_\mathrm{CB^\#}
\end{equation}

\noindent For the fitting of the relative photoproduct populations, we considered a delayed kinetics where the reaction starts at time $t_0$ after excitation. The population of excited cyclobutanone was fitted to 

\begin{equation}
    p_{\mathrm{CB}}(t)=
    \begin{cases}
        1 & \text{if } t \in [0, t_0]\\
        \mathrm{e}^{-k_\mathrm{CB}(t-t_0)} & \text{if } t > t_0
    \end{cases}
\end{equation}

\noindent where the delay time $t_0$ and the rate constant $k_\mathrm{CB}$ are the fitting parameters. The total lifetime of the excited cyclobutanone was then evaluated as $\tau_{\mathrm{CB}} = t_0 + 1/k_\mathrm{CB}$, i.e., the time it takes for the cyclobutanone population to reach a value of $1/e$. The relative population of the open structure was fitted to the following function

\begin{equation}
    p_{\mathrm{open}}(t)=
    \begin{cases}
        0 & \text{if } t \in [0, t_0]\\
        \frac{k_\mathrm{CB}}{k_\mathrm{open}-k_\mathrm{CB}} \left( \mathrm{e}^{-k_\mathrm{CB}(t-t_0)} - \mathrm{e}^{-k_\mathrm{open}(t-t_0)} \right) & \text{if } t > t_0
    \end{cases}
\end{equation}

\noindent where $k_\mathrm{open}$ was the only fitted parameter. The lifetime was then obtained analogously as $\tau_\mathrm{open}=1/k_\mathrm{open}$. Finally, the relative populations of photoproducts were fitted to 

\begin{equation}
    p_{i}(t)=
    \begin{cases}
        0 & \text{if } t \in [0, t_0]\\
        k_\mathrm{i}\left[\frac{k_\mathrm{CB}}{k_\mathrm{CB}-k_\mathrm{open}} \left( \frac{\mathrm{e}^{-k_\mathrm{CB}(t-t_0)}}{k_\mathrm{CB}} - \frac{\mathrm{e}^{-k_\mathrm{open}(t-t_0)}}{k_\mathrm{open}} \right) + \frac{1}{k_\mathrm{open}}\right] & \text{if } t > t_0
    \end{cases}
\end{equation}

\noindent where $k_i$ was the fitting parameter, $i$ standing for CO, EK, or CB$^\#$. The rate constants $k_\mathrm{i}$ were finally inverted to provide lifetimes of their respective reactions. The FSSH relative photoproduct populations, the fit of the photoproduct populations, and the lifetimes of cyclobutanone and the open structure are provided in Fig.~\ref{fig:fssh_fit}. The lifetimes of the photoproducts depend solely on the ratio of their populations and are $\tau_\mathrm{CO}=189$ fs, $\tau_\mathrm{EK}=329$ fs and $\tau_\mathrm{CB^\#}=2368$ fs.

\clearpage
\section{Comparing LZSH and FSSH}

\begin{figure}[ht]
    \centering
    \includegraphics[width=1.0\textwidth]{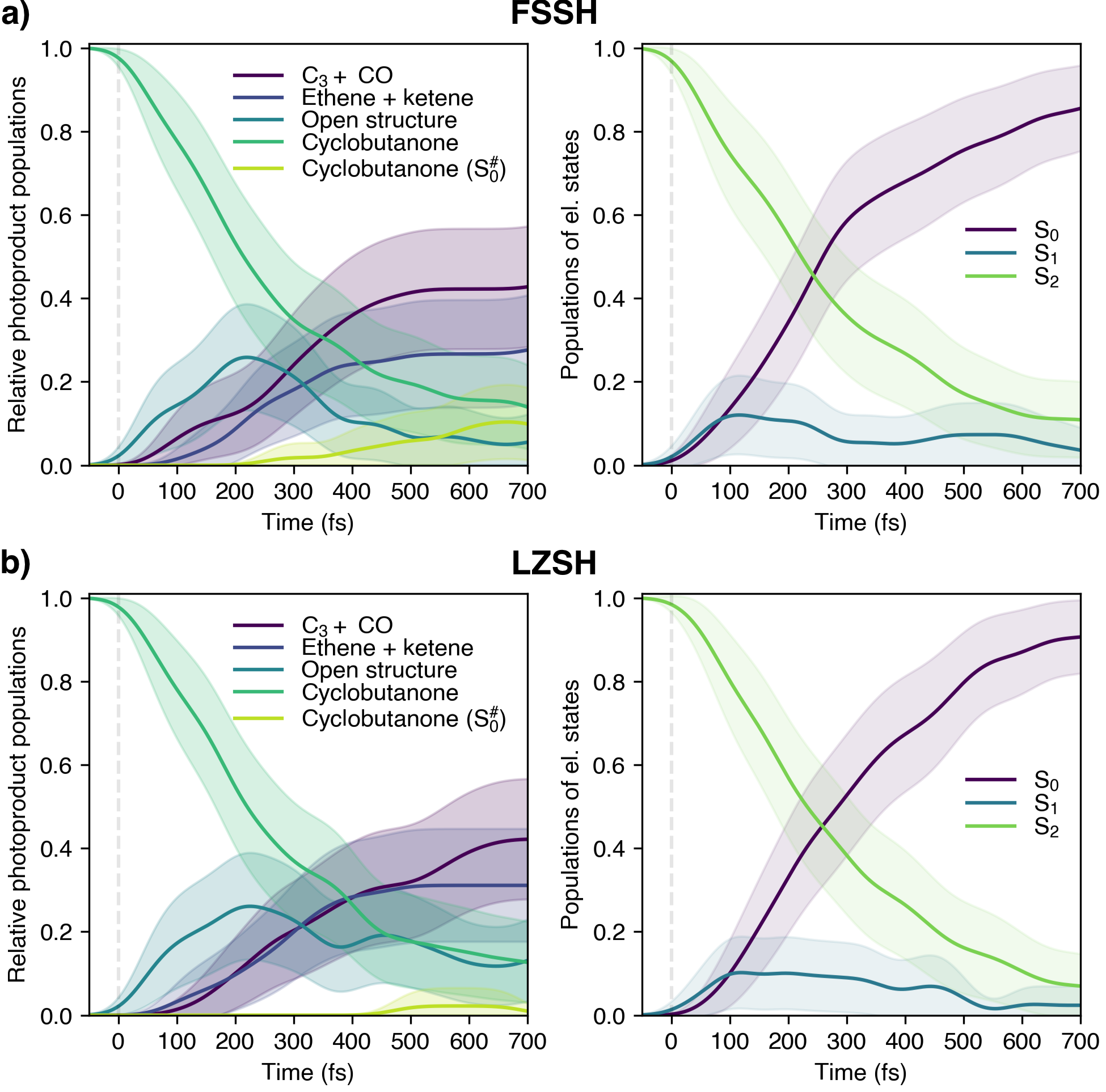}
    \caption{Relative photoproduct and electronic-state populations predicted by a) FSSH and b) LZSH for the same set of 45 initial conditions and 700~fs of dynamics. XMS-CASPT2 was used for both methods (with the same active space and basis set).}
    \label{fig:comparison}
\end{figure}

\clearpage
%
%